\documentclass[%
 reprint,
nofootinbib,
 amsmath,amssymb,
 aps,
floatfix,
]{revtex4-2}

\usepackage{lipsum}

\usepackage{graphicx}
\usepackage{dcolumn}
\usepackage{bm}
\usepackage{float}

\usepackage{color}
\usepackage[dvipsnames]{xcolor}
\definecolor{blueprl}{RGB}{46,48,146}

\usepackage[breaklinks=true]{hyperref}
\hypersetup{
colorlinks   = true, 
urlcolor     = blueprl, 
linkcolor    = blueprl, 
citecolor    = blueprl 
}

\usepackage[FIGBOTCAP]{subfigure}

\usepackage{cleveref}
\crefname{equation}{Eq.}{Eqs.}
\Crefname{equation}{Equation}{Equations}
\crefname{figure}{Fig.}{Figs.}
\Crefname{figure}{Figure}{Figures}
\crefname{figure}{Fig.}{Figs.}
\Crefname{figure}{Figure}{Figures}
\crefname{section}{Supplemental Material Section}{Supplemental Material Sections}
\Crefname{section}{Supplemental Material Section}{Supplemental Material Sections}
\crefname{appendix}{Appendix}{Appendices}
\Crefname{appendix}{Appendix}{Appendices}
\crefname{table}{Table}{Tables}
\Crefname{table}{Table}{Tables}

\newcommand{\bra}[1]{\mbox{$\langle #1 |$}}
\newcommand{\ket}[1]{\mbox{$| #1 \rangle$}}

\def\braket#1#2{{\langle#1\vert#2\rangle}}

\usepackage{enumitem,amssymb}
\newlist{todolist}{itemize}{2}
\setlist[todolist]{label=$\square$}
\usepackage{pifont}

\usepackage{etoolbox}

\begin{document}

\title{Deterministic preparation of optical squeezed cat and Gottesman-Kitaev-Preskill states}

\author{Matthew S. Winnel}\email{mattwinnel@gmail.com}
\affiliation{Centre for Quantum Computation and Communication Technology, School of Mathematics and Physics, University of Queensland, St Lucia, Queensland 4072, Australia}
\author{Joshua J. Guanzon} 
\affiliation{Centre for Quantum Computation and Communication Technology, School of Mathematics and Physics, University of Queensland, St Lucia, Queensland 4072, Australia}
\author{Deepesh Singh}
\affiliation{Centre for Quantum Computation and Communication Technology, School of Mathematics and Physics, University of Queensland, St Lucia, Queensland 4072, Australia}
\author{Timothy C. Ralph}
\affiliation{Centre for Quantum Computation and Communication Technology, School of Mathematics and Physics, University of Queensland, St Lucia, Queensland 4072, Australia}

\date{\today}

\begin{abstract}
Large-amplitude squeezed cat and high-quality Gottesman-Kitaev-Preskill (GKP) states are powerful resources for quantum error correction. However, previous schemes in optics are limited to low success probabilities, small amplitudes, and low squeezing. We overcome these limitations and present scalable schemes in optics for the deterministic preparation of large-amplitude squeezed cat states using only Gaussian operations and photon-number measurements. These states can be bred to prepare high-quality approximate GKP states, showing that GKP error correction in optics is technically feasible in near-term experiments.
\end{abstract}

\maketitle

{\it Introduction.} Before quantum technologies can provide significant advantages in the real world~\cite{Arute_2019,RevModPhys.81.1301}, they must overcome noise. Quantum error correction~\cite{Gottesman_error_correction} provides one possible solution.

Encodings of quantum information can be characterised as discrete variable (DV) or continuous variable (CV)~\cite{Braunstein_2005,yonezawa2008continuousvariable,cerf2007quantum,Weedbrook_2012,Serafini2017QuantumCV}. A CV state can cleverly exploit the infinite-dimensional Hilbert space of oscillators to encode a DV system, offering protection against noise for quantum information processing. Notable examples are cat~\cite{Mirrahimi_2014,PhysRevA.94.042332,PhysRevLett.119.030502}, squeezed cat~\cite{Xu2023,PhysRevA.106.022431,PhysRevLett.120.073603} and Gottesman-Kitaev-Preskill (GKP)~\cite{GKP2001} states.

However, large-amplitude cat and high-quality GKP states are difficult to realize in optics~\cite{doi:10.1063/1.5025456,Sychev_2017}. Previously, the practical method to generate cat states relied on Gaussian squeezed-vacuum states and conditional photon-number measurements on a beamsplitter (photon subtraction)~\cite{PhysRevA.55.3184}. Despite promising experimental progress~\cite{PhysRevLett.114.193602,Endo_2023}, the shortcomings are always that the prepared states have low amplitude, low squeezing, or are highly non-deterministic~\cite{PhysRevA.106.043721,
PhysRevX.13.031001,Ourjoumtsev2007,PhysRevA.103.013710,PhysRevA.82.031802,https://doi.org/10.48550/arxiv.2212.05436,https://doi.org/10.48550/arxiv.2212.08827,Ourjoumtsev2007,Eaton_2022}.

In this paper, we solve these limitations and show how to prepare large-amplitude squeezed cat states with high probability of success. We present two schemes. The first scheme deterministically converts a source of large Fock states into large-amplitude squeezed cat states. While there has been recent progress in the deterministic generation of large Fock states~\cite{Xia:12,PhysRevLett.125.093603} and photon-number-resolving detection~\cite{Provaznik:20,Gerrits:12,Lita:08,Fukuda:11,Sridhar:14}, this remains challenging. Hence, we introduce a second scheme for deterministically preparing cat states, and by introducing some non-determinism, large-amplitude cat states can be ensured at the output for a smaller Fock resource state, or arbitrary even/odd photon-number parity states can be used as input.

Optical GKP states are also difficult to realize. A promising method to non-deterministically prepare GKP states is via ``Gaussian boson sampling''-like devices~\cite{Bourassa_2021,sabapathy2019production,su2019conversion,quesada2019simulating,tzitrin2020progress}. GKP states can also be prepared using cavity QED~\cite{PhysRevLett.128.170503}, and there are schemes based on photon catalysis to make cat and GKP states~\cite{PhysRevA.86.043820,Eaton_2019}. The simplest method for deterministically preparing GKP states is by breeding large-amplitude squeezed cat states~\cite{Vasconcelos:10,PhysRevA.97.022341, Policarpo2016}. Thus, our schemes solve the GKP preparation problem and universal fault-tolerant quantum computation is achievable using our schemes for state preparation~\cite{PhysRevLett.123.200502}.

Ref.~\cite{konno2023propagating} recently demonstrated propagating GKP states in optics. Their scheme can be seen as the non-deterministic lowest order of our scheme, which gives strong credence that our proposal is technically feasible in the near-term.

We provide numerical simulations to validate our approximate analytical results. In~\cite{Matt_code}, we provide our MATLAB code~\cite{MATLAB} and our python code~\cite{python} which perform simulations in the Fock basis. To perform numerical calculations, we truncate the infinite-dimensional Fock space and ensure it is high enough so our results are valid. 

{\it Preliminaries.} Our aim is to deterministically prepare squeezed cats of the form:
\begin{align}
\ket{\mathcal{C}_{\alpha,r}^{\pm}} &\equiv \mathcal{N} ( \ket{\alpha,r} \pm \ket{-\alpha,r}),\label{eq:cat}
\end{align}
where $\ket{\alpha,r} \equiv \hat{D}(\alpha)\hat{S}(r)\ket{0}$ is a displaced squeezed vacuum state in the position-quadrature direction, $\alpha$ is real, and $\mathcal{N}$ is the normalisation constant. The $+$ and $-$ superpositions consist of only even or odd photon numbers, respectively. The displacement and squeezing operators are defined as $\hat{D}(\alpha)\equiv e^{\alpha \hat{a}^\dagger - \alpha^* \hat{a}}$ and $\hat{S}(r,\theta) \equiv  e^{\frac{r}{2}(\hat{a}^2e^{-2i\theta}-\hat{a}^{\dagger2}e^{2i\theta})}$, respectively. For squeezing ($r>0$) and anti-squeezing ($r<0$) we write $\hat{S}(r)\equiv \hat{S}(r,\theta=0)$. The phase rotation operator is $\hat{R}(\theta) = e^{i\theta\hat{a}^\dagger\hat{a}}$ where $\theta$ is the rotation angle in radians. Gaussian operations can transform squeezed cats into the form of~\cref{eq:cat}. We use the following conventions for position $\hat{q}=(\hat{a}+\hat{a}^\dagger)/{2}$ and momentum $\hat{p} = i(\hat{a}^\dagger - \hat{a})/{2}$, in natural units of $\hbar =1/2$.

Our schemes prepare states that are qualitatively similar to~\cref{eq:cat} though they may not be exactly squeezed cats. Imperfections away from~\cref{eq:cat} can be made arbitrarily small for large amplitude and are ``squeezed out'' during GKP breeding. Furthermore, imperfections can be fixed during GKP breeding via GKP error correction thanks to the error-correcting properties of the GKP code. It is also known that squeezed cats enjoy higher tolerance to losses compared to standard cats~\cite{PhysRevLett.120.073603}.

The ideal infinite-energy square GKP states~\cite{GKP2001} can be written as superpositions of quadrature eigenstates. One possibility for constructing physical GKP states is by replacing the position eigenstates with displaced squeezed vacuum and applying a Gaussian envelope, allowing for physical square GKP computational basis states to be defined as~\cite{PhysRevA.97.032346}
\begin{gather}
    \ket{0_\text{GKP}^\Delta} \propto  \sum_{u \in\mathbb{Z} } e^{-\frac{\pi}{2}\Delta^2 (2u)^2 }  \ket{2u\sqrt{\frac{\pi}{2}},-\ln \Delta}, \label{eq:physicalgpk0} \\
    \ket{1_\text{GKP}^\Delta} \propto  \sum_{u \in\mathbb{Z} } e^{-\frac{\pi}{2}\Delta^2 (2u+1)^2 }  \ket{(2u+1)\sqrt{\frac{\pi}{2}},-\ln \Delta}, \label{eq:physicalgpk1}
\end{gather}
where $\Delta \in (0,1]$ gives $-10 \log_{10}(\Delta^2)$ dB of GKP squeezing, tending towards ideal GKP states as $\Delta\to0$.

Squeezed cats serve as approximate GKP states, in particular: 
\begin{align}
 \ket{1_\text{GKP}^{\Delta}}  &\propto \ket{{-}\sqrt{\frac{\pi}{2}},{-}\ln \Delta} + \ket{\sqrt{\frac{\pi}{2}},{-}\ln \Delta}\label{eq:target_state},
\end{align}
which is \cref{eq:physicalgpk1} except with only the two most important central components $u\in\{-1,0\}$.

\begin{figure}
\centering
\includegraphics[width=1\linewidth]{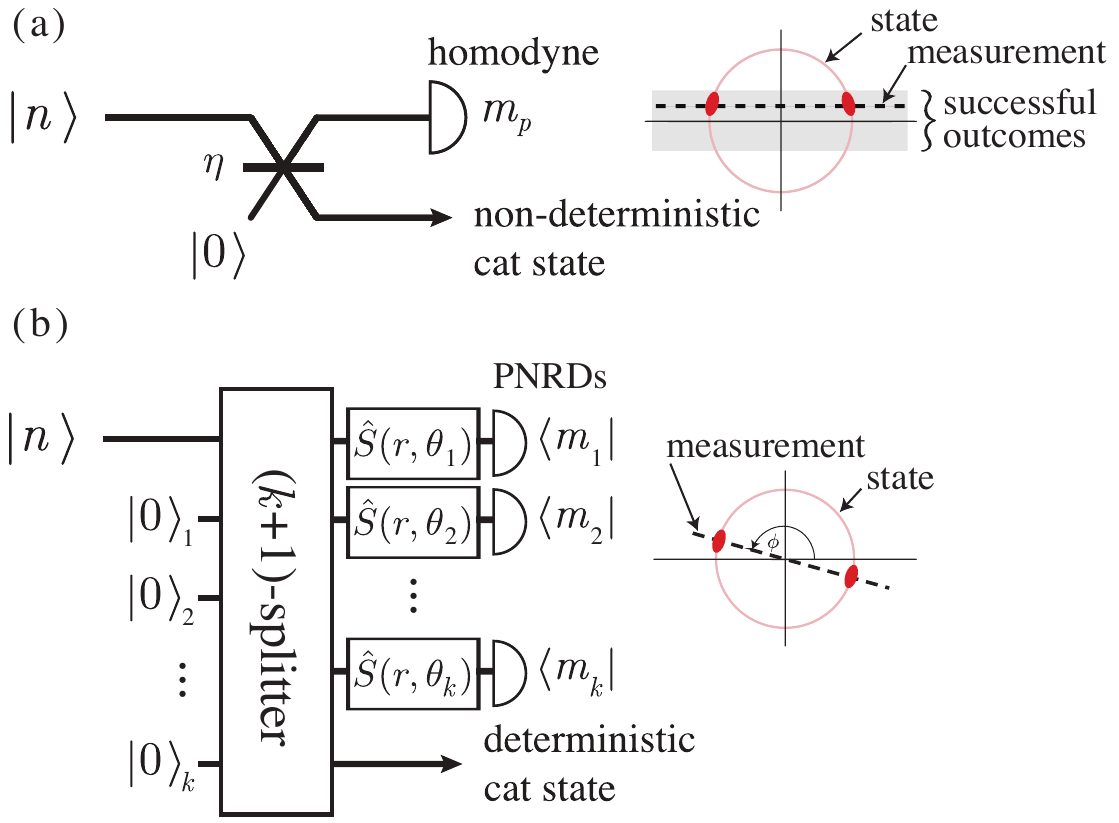}
 \caption{Optical circuits for the preparation of squeezed cat states from Fock photon-number states. In (a), we present the non-deterministic scheme from Ref.~\cite{Ourjoumtsev2007}. A Fock state $\ket{n}$ is mixed with vacuum on a beamsplitter with transmissivity $\eta=1/2$ and homodyne detection is performed on one of the output modes. Squeezed cat states are prepared on the other mode so long as the homodyne measurement outcome is close to zero, as shown in the phase space illustration. In (b), we present our scheme I which is deterministic. We split the Fock state via a $(k+1)$-splitter. The light is divided essentially evenly between the first $k$ modes, then squeezed in different directions $\theta_j$, before being measured via photon-number-resolving detectors (PNRDs). A squeezed cat state is deterministically prepared on the remaining mode. The squeezing and amplitude of the prepared cat state be tuned via the ($k+1$)-splitter and the size of $n$. For $k>2$ we can choose $\theta_j = j\pi/k$ where $j = 1,2,...,k$ and for $k=2$ we can choose $\theta_j = j\pi/4$.}\label{fig:new_figure}
\end{figure}

{\it Scheme I.} Ref.~\cite{Ourjoumtsev2007} introduced a non-deterministic scheme given in~\cref{fig:new_figure}(a), which consists of mixing an $n$ Fock state with vacuum and then performing homodyne detection on one quadrature $p$. By post-selecting on or near the zero outcome $m_p\sim0$, an approximate {\it{squeezed}} 2-cat is produced. For intermediate $m_p$ outcomes the cat components are not parallel and for large $m_p$ outcomes it is not a cat state. Hence it is the homodyne detection which limits Ref.~\cite{Ourjoumtsev2007}, thus a different type of measurement is required to achieve determinism.

Our scheme I in~\cref{fig:new_figure}(b) solves this problem, and therefore can deterministically prepare squeezed cats from Fock states. The Fock state is split into $k+1$ modes using a $(k+1)$-splitter, so that roughly equal amounts propagate towards each of the $k$ detectors with the remaining amount becoming the prepared cat state on the final mode. Each of the first $k$ modes is squeezed in different directions before being measured by photon-number-resolving detection (PNRD). For large $n$, two-component squeezed cat states are deterministically prepared in the remaining mode with high fidelity. This detection scheme acts like an effective phase measurement which runs through the origin of phase space.

Overall, the Fock state experiences a total transmissivity of $\eta_{\text{total}}$ so the amplitude of the prepared cat is roughly $|\alpha| \approx \sqrt{n\eta_{\text{total}}}$. The cat state is also squeezed with squeezing parameter roughly $r \approx -0.5\ln(1-\eta_\text{total})$. That is, by using an input Fock state, all prepared cats have roughly the same amplitude and are positively squeezed. These are great advantages for breeding GKP states from cats.

The squeezing on each mode is performed in different directions $\theta_j$ to prevent the possibility of preparing four-component cats. To do this, we must have at least $k\geq 2$. Then, for $k>2$, we can choose the squeezing angles to be equally distant around the circle, i.e., $\theta_j = j\pi/k$ where $j=1,2,...,k$. For $k=2$, since there are only two angles we better choose for example $\theta_j = j\pi/4$ (choosing $\theta_j = j\pi/2$ sometimes prepares 4-cats). For large $k$, it may be more practical to randomly choose $\theta_j$ which works well.

The prepared squeezed cat depends on the following experimental parameters: the Fock number $n$, the number of detected modes $k$, and on the squeezing directions $\theta_j$. The prepared cat also depends on the outcomes at the detectors $m_j$. However, the scheme is deterministic since for any possible series of outcomes the output is a squeezed cat.

Interestingly, the non-deterministic scheme from Ref.~\cite{Ourjoumtsev2007} shown in~\cref{fig:new_figure}(a) is equivalent to  our scheme I shown in~\cref{fig:new_figure}(b) for the following very specific choice of our parameters: $k=1$, $r\to\infty$, $\theta_1 = \pi/2$, $\eta_\text{total} = \eta$, and we post-select on outcome $m_1=0$.

To prove scheme I works, we expand the $k$-splitter in time and investigate the effect of the squeezing and PNRD measurement outcomes on the evolving cat. Later, to make it more practical, we introduce scheme II which does not require inline squeezing and prepares higher-fidelity cat states.

Consider scheme I for $k=1$. An $n$ Fock state is mixed with vacuum on a beamsplitter, then on one output mode squeezing followed by PNRD is performed. This determistically prepares approximate squeezed cats but with two or four components depending on the outcome at the detector in the other mode. These squeezed cats are ideal with unit fidelity for large $n$. This is shown in the following circuit~\cite{SUPP}: \par
\makebox[0.925\linewidth][c]{\includegraphics[width=0.9\linewidth]{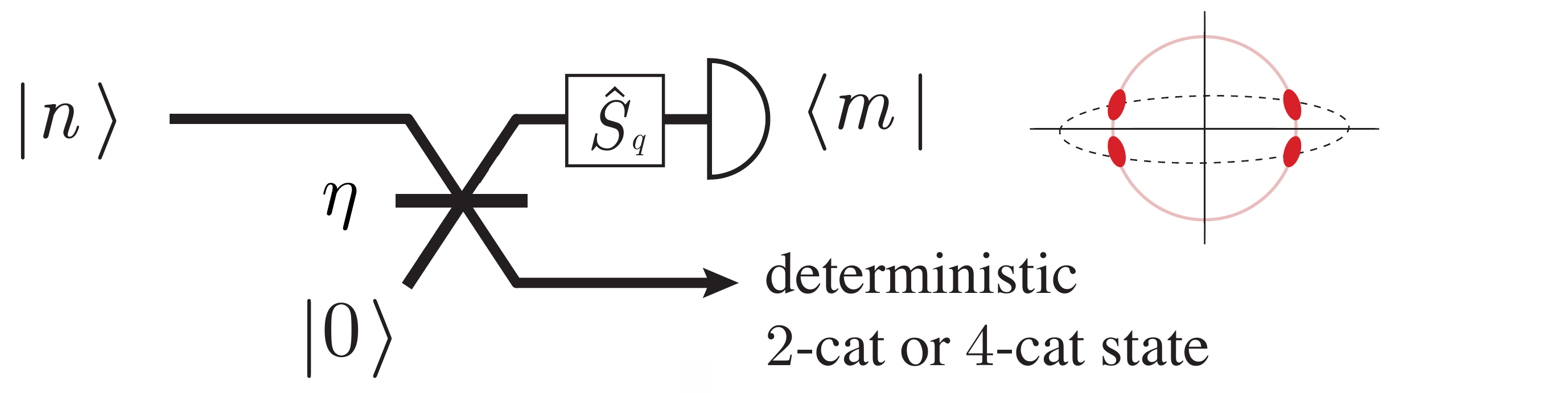}}
\noindent

\begin{figure}
\centering
\includegraphics[width=1\linewidth]{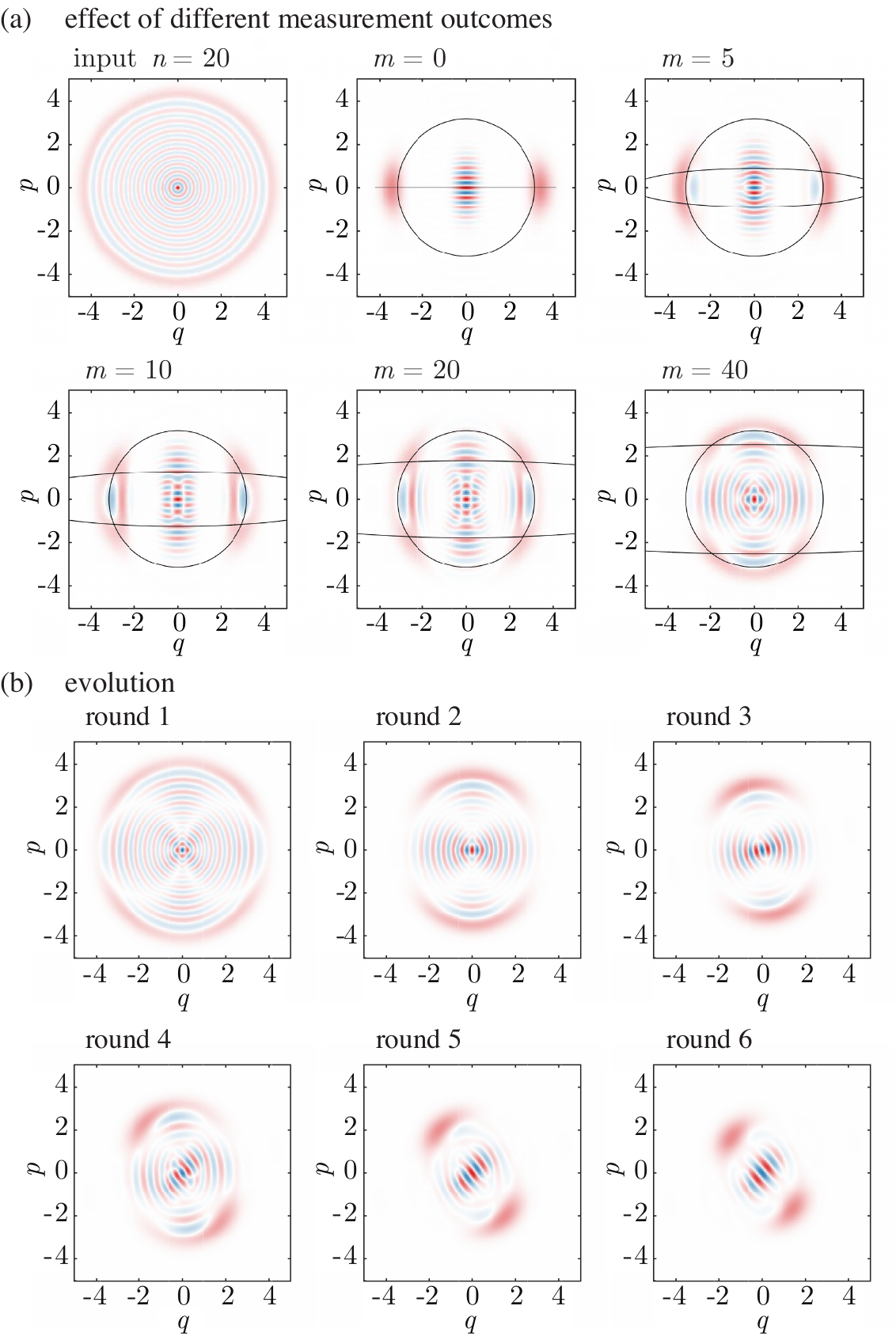}
 \caption{Numerical results for deterministic cat-state preparation using scheme I with input Fock state with $n=20$ and 8 dB of squeezing before the PNRDs. (a) Wigner functions plot of the input state and output states for different outcomes $m$ for the first iteration of scheme I with $\eta=1/2$. The components lie approximately at the intersection of a circle (input) and an ellipse (measurement) as illustrated by solid lines in the figure. (b) Wigner function plots of the evolving cat at round $j$, with total transmissivity $\eta^k=1/4$, and total number of rounds $k = 6$. The final cat is positively squeezed since the Fock state provides high certainty in the absolute amplitude.}\label{fig:cat_prep_scheme_I}
\end{figure}

In the limit of large inline squeezing, the components of the prepared cat are located at the intersections of the circle $n \eta = |\alpha|^2 $ and the ellipse $b = m \eta /(1-\eta)  =  e^{2r}	\text{Re}(\alpha)^2  + e^{-2r} \text{Im}(\alpha)^2$. Wigner function plots of the initial state and the prepared cats for different measurement outcomes are plotted in~\cref{fig:cat_prep_scheme_I}(a) for $n=20$ and 8 dB of inline squeezing before the PNRD with transmissivity $\eta=1/2$.

The initial measurement projects onto a superposition of four angles $\{\phi_1,-\phi_1, \phi_1+\pi, -\phi_1+\pi\}$ where $|\phi_1| \in [0, \pi)$. To deterministically prepare 2-cats we simply repeat the circuit, squeezing in a rotated direction relative to the state to break the reflection symmetry. This selects one pair of components out of the initial four. As $n$ becomes very large, the components are totally distinguishable, proving our scheme deterministically prepares ideal squeezed cats with unit fidelity for large $n$.

For finite $n$, the scheme prepares squeezed 2-cats with very high probability and very high fidelity, as long as $n$ is large enough and $\eta$ small enough so that the components of the prepared cat are well distinguishable. A proof is given in~\cite{SUPP}. It can be advantageous to repeat the protocol $k>2$ times, squeezing in different directions (or rotating the state by different amounts), thus, smoothing out the evolution towards a cat and effectively reproducing the evolution of~\cref{fig:new_figure}(b). This is summarised in the following circuit:\par
\makebox[0.925\linewidth][c]{\includegraphics[width=1\linewidth]{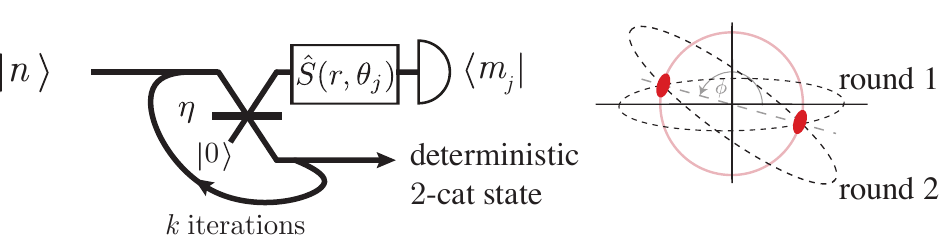}}
\noindent Since the input Fock state is being attenuated, the transmissivity of the beamsplitters $\eta_j$ should decrease with $j$ in order to ensure equal amounts of light are measured by each detector. However, this is not essential and it is more practical to have a constant transmissivity $\eta_j = \eta$ chosen so that the desired total transmissivity is $\eta_{\text{total}} = \eta^k$. This is the situation we consider from now on. We give details in~\cite{SUPP}.

The squeezing angle $\theta_j \in [0, \pi)$ can be chosen randomly. Although no feed-forward is necessary, we could exploit the knowledge of the component locations after the first round in order to maximise the probability that they can be distinguished in following rounds. We leave exploration of this possibility for future work.

The cat components have amplitude $|\alpha|\approx \sqrt{n\eta^k}$ and squeezing  $r\approx-0.5\ln(1-\eta^k)$, and the number parity is $n+\sum_j m_j$, where $m_j$ is the detected number of photons at round $j$.

In~\cref{fig:cat_prep_scheme_I}(b), we show an example of how the initial Fock state evolves deterministically towards a 2-cat using scheme I where $n=20$, $k=6$, $\eta^k=1/4$, and there is 8 dB of inline squeezing. 

{\it Scheme II.} We introduce scheme II to overcome some limitations of scheme I. We first note that the phase probability distribution of an arbitrary input state evolves the same way in the following two circuits in the limit of high transmissivity $\eta$:\par
\makebox[0.925\linewidth][c]{\includegraphics[width=1\linewidth]{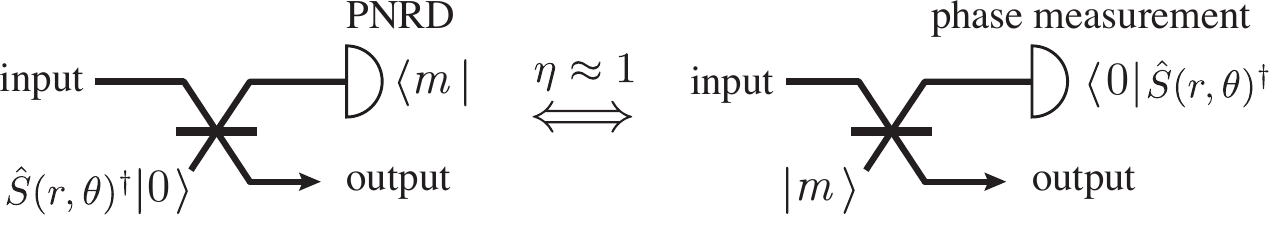}}
\noindent We do not have a phase measurement required by the right circuit, but we do have squeezed vacuum required by the left circuit. Scheme II works by iterating this circuit as to effectively perform a phase measurement and to deterministically prepare cats:\par
\makebox[0.925\linewidth][c]{\includegraphics[width=1\linewidth]{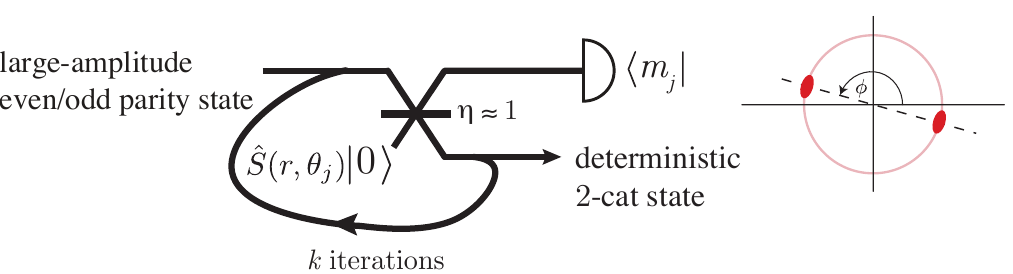}}
\noindent The squeezing angle $\theta_j \in [0, \pi)$ should be different each round. The parity of the output cat is given by the parity of $\sum m_j$ + $\mu$, where $\mu$ is the parity of the initial state and $m_j$ is the detected number of photons at round $j$. We show in~\cite{SUPP} that the phase probability distribution approaches two peaks at $\phi$ and $\phi+\pi$ where $\phi \in(0,\pi]$.

The input state to scheme II can be any arbitrary state with at least 2-fold rotation symmetry (i.e., even or odd Fock-number parity). Therefore, squeezed Fock states in particular can be used as an input. Squeezing can be used to greatly reduce the size requirement of the Fock number resource.

The small-amplitude cats can non-deterministically be removed with high probability for sufficiently-large amounts of squeezing. That is, squeezing is a powerful resource for increasing the achievable amplitude and mean photon number of the prepared cats for a smaller $n$. The mean photon number of a squeezed Fock state $\hat{S}(r)\ket{n}$ is $\bar{n}_n(r) = (2n+1) \sinh^2{r} + n$~\cite{dantas_1998}. The series expansion at $r=0$ is $n+(2n+1)r^2$ which only increases linearly with $n$ but for large $r$ the mean photon number increases exponentially as $e^{2r}$.  Despite the advantages of increasing squeezing, it is always desirable to increase $n$ as well if possible. In particular, it is much more desirable to use a squeezed single photon input compared with a squeezed vacuum input since, for small amounts of squeezing, the former already contains a photon whereas the later contains only $\sim r^2$ photons. This shows how our scheme II, non-deterministically removing the small-amplitude cats and using squeezed single photons or squeezed Fock states as the input resource states, immediately out-performs the non-deterministic cat preparation protocols based on photon subtraction from only vacuum resource states~\cite{PhysRevA.55.3184,PhysRevA.103.013710,PhysRevA.82.031802,https://doi.org/10.48550/arxiv.2212.08827}. Plots and further details on the trade-off between the success probability and the obtainable size of the non-deterministic cats using a source of squeezed Fock states is presented in~\cite{SUPP}.

Another advantage is that scheme II requires no inline squeezing before the detector.

\begin{figure}
\centering
 \includegraphics[width=1\linewidth]{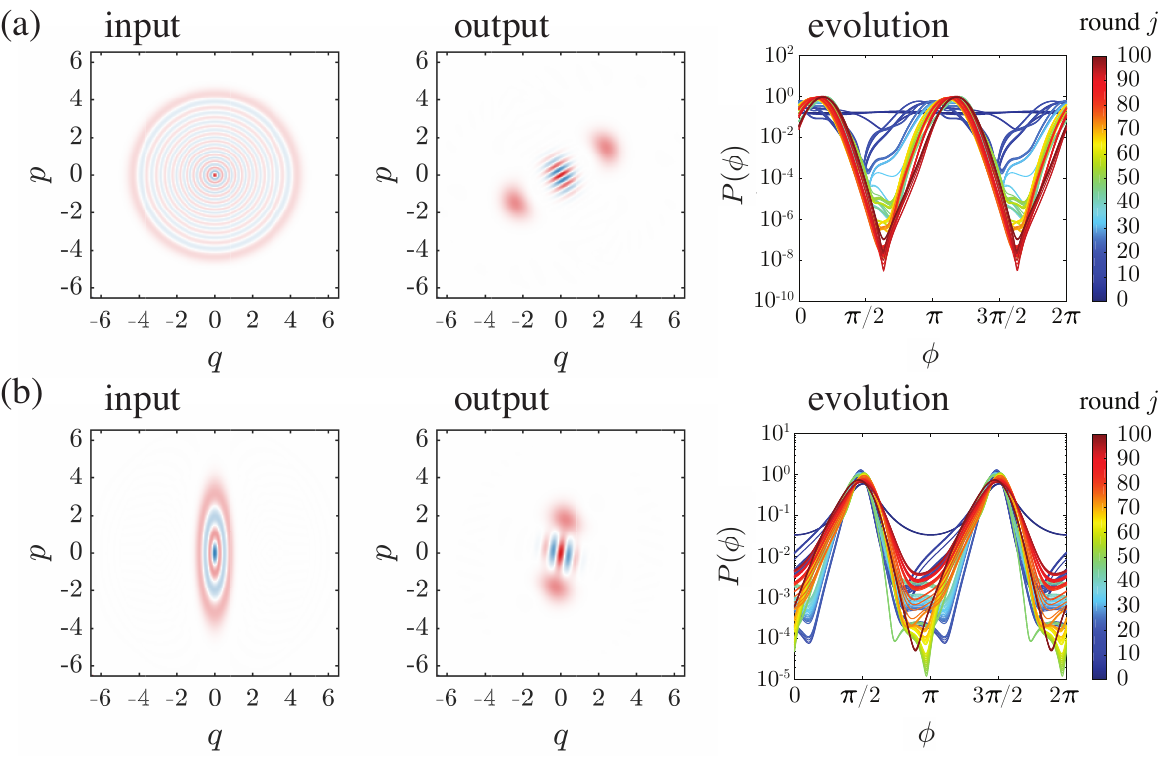}
 \caption{Simulation results for scheme II for the deterministic preparation of squeezed two-component cat states from large-amplitude even/odd parity states, with total transmissivity $\eta^k=1/2$, total number of rounds $k=100$, and squeezed vacuum with 6 dB of squeezing. We show results for different input states: (a) input Fock state $|n\rangle$ with $n=20$ and (b) input squeezed Fock state $\hat{S}(r)\ket{n}$ with $n=3$ and 6 dB of squeezing. We show the Wigner functions of the input state (left), output state (middle), and the evolution (from blue to red) of the phase probability distribution (right). }\label{fig:cat_prep_scheme_II}
\end{figure}

Numerical simulations of scheme II are shown for a Fock state input $\ket{20}$ in~\cref{fig:cat_prep_scheme_II}(a), and for a squeezed Fock state input $\hat{S}(r)\ket{3}$ with 6 dB of squeezing in~\cref{fig:cat_prep_scheme_II}(b). The total transmissivity is $\eta^k=1/2$, the squeezed vacuum has 6 dB of squeezing, and there are $k=100$ rounds. This shows the asymptotic behaviour of the protocol for large $k$. Note that $k$ does not need to be this large in practice. We track the phase of our state using the phase probability distribution $P(\phi)$, as defined in Sec. II A of Ref.~\cite{Wiseman1997AdaptiveSP}. In these figures, we plot the Wigner functions of the initial state (left), final state (center), and we plot the evolution of the phase probability distribution $P(\phi)$ (right).

{\it Fidelity and squeezing.} To showcase the quality of our squeezed cats, we compute the fidelity and squeezing. We first correct the prepared cat using phase rotations and squeezing so that the components of the cat have mean values at $\alpha = \pm \sqrt{\pi/2}$ which is the square GKP grid spacing. The expected squeezing of the corrected prepared cat is given by~\cite{SUPP}
\begin{align}
    \Delta = \sqrt{{\pi (1-\eta^k)}/{(2n \eta^k})}.\label{eq:expected_squeezing}
\end{align}

\begin{figure}
\centering
 \includegraphics[width=1\linewidth]{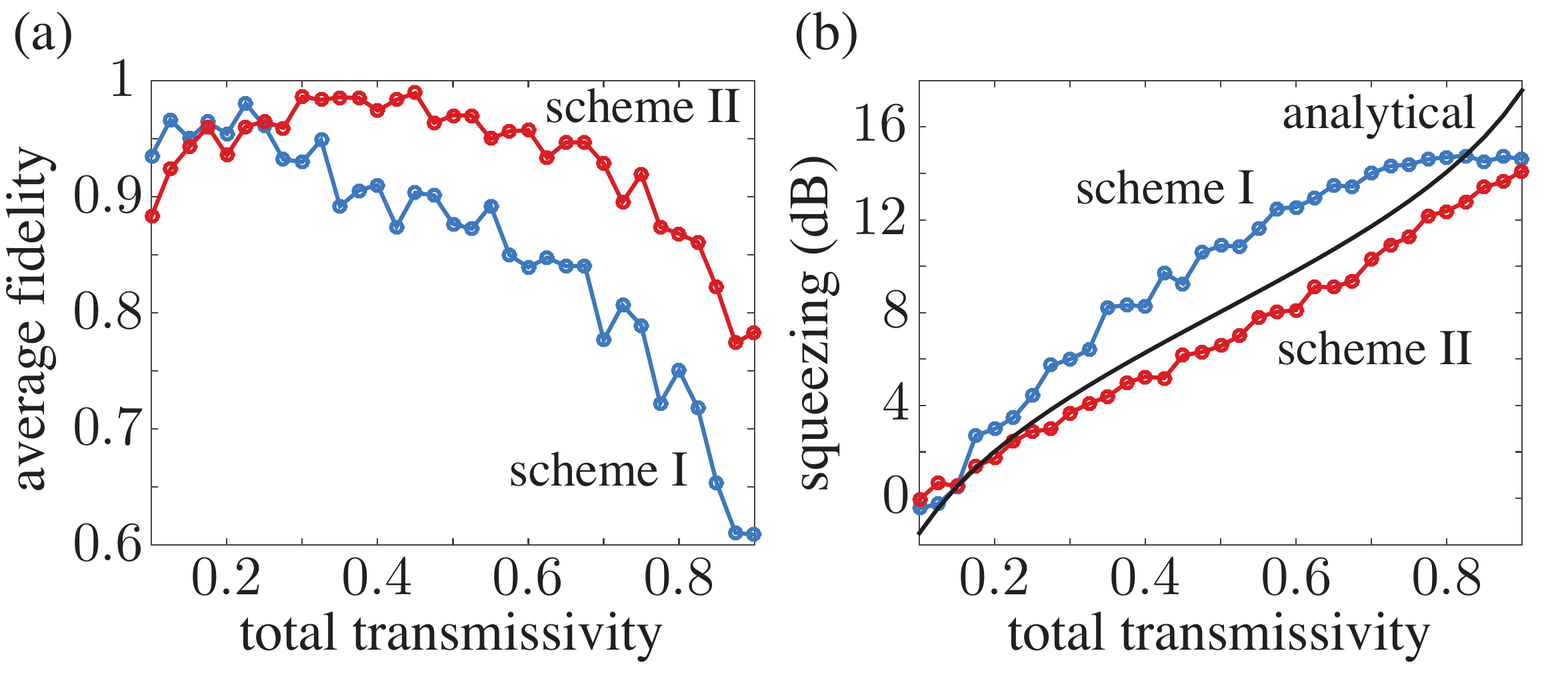}
 \caption{Quality of the prepared squeezed cats using scheme I (blue) or scheme II (red) with $n=10$. We average over 50 runs of the simulation and show in (a) the average fidelity and in (b) the average squeezing (computed numerically from the variance), where, for scheme I, $k=6$ and there is 6 dB of inline squeezing, and for scheme II, $k=100$, and the squeezed vacuum has 6 dB of squeezing. In (b), we also plot the analytical expected squeezing (black) as given by~\cref{eq:expected_squeezing}. We find a trade-off between fidelity and squeezing for a given $n$.}\label{fig:fidelity}
\end{figure}

In~\cref{fig:fidelity}, we plot the average fidelity of the prepared cats with the target cat in (a) and average cat squeezing in (b) averaging over 50 random runs of the experiment for scheme I (blue) and scheme II (red) as a function of total transmissivity $\eta^k$ for a fixed Fock state with $n=10$. The target state has an amount of squeezing determined from the variance in the $p$ quadrature of the prepared cat. In (b), we also plot the expected squeezing~\cref{eq:expected_squeezing} (black). For scheme I, the total number of rounds is $k=6$ and 6 dB of inline squeezing. For scheme II, the total number of rounds is $k=100$ and the squeezed vacuum has 6 dB of squeezing. We find a trade-off between fidelity and squeezing. These average results can be improved by post-selecting only the best cats.

\begin{figure}
\centering
 \includegraphics[width=1\linewidth]{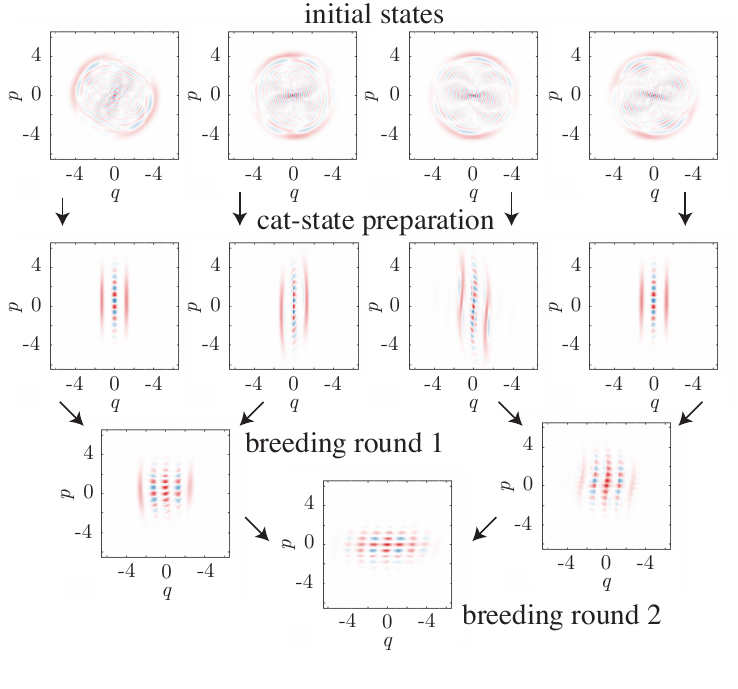}
 \caption{Deterministically breeding GKP states~\cite{80641,Vasconcelos:10,PhysRevA.97.022341} from our deterministic source of squeezed cats prepared using scheme II via a source of arbitrary large-amplitude input states of the form $\propto \sum_{n=0}^{10} c_n 1.2^{2n} \ket{2n}$ where $c_n$ are arbitrary random numbers, with larger weight on the larger Fock numbers given by $1.2^{2n}$. We show two rounds of breeding via GKP-CNOT gates and homodyne detection of the $p$ quadrature. The states are prepared and combined randomly, simulating an experiment, as indicated by the arrows.}\label{fig:GKP_breeding_geometry}
\end{figure}

{\it GKP error correction.} GKP states can be bred deterministically by entangling two primitive GKP qubits with GKP-CNOT gates and measuring one output mode with homodyne detection~\cite{80641,Vasconcelos:10,PhysRevA.97.022341}. States are first corrected into the form given by~\cref{eq:target_state}. Odd cats can be corrected via a small displacement in the $p$ direction so that the Wigner function has a local maximum at the origin. Numerical results breeding GKP from arbitrary even number-parity states are shown in~\cref{fig:GKP_breeding_geometry}.

\Cref{eq:expected_squeezing} relates $n$ with $\Delta$ for fixed total transmissivity $\eta^k$. Setting $\eta^k = 1/2$, we see that for large $n$ the amount of squeezing is very large and the fidelity is very high, i.e., $\lim_{n\to\infty} \Delta = 0$ and $\lim_{n\to\infty} F = 1$. The fidelity is high for large $n$ because the effective phase measurements during cat-state preparation have vanishing phase uncertainty. Then, approximate GKP states can be prepared with very high quality (in terms of fidelity and squeezing) and fault-tolerant quantum computation using GKP error correction is possible using our schemes for state preparation.

We consider the performance of our states for GKP error correction in~\cite{SUPP}. To reduce the required $n$, we can have $\eta^k>1/2$, but there is a trade-off between squeezing and fidelity as shown in~\cref{fig:fidelity}. To reduce the required $n$ further, we can allow some non-determinism. 

It is challenging to implement inline squeezing experimentally which is required to correct the large-amplitude cats onto the GKP lattice for breeding. Increasing the squeezing of the prepared cat and GKP states may be achieved by combining the states with squeezed vacuum via a beamsplitter or GKP-CNOT gate and measuring one output mode with a homodyne detector. A similar procedure can be used combining cats and non-deterministically increasing the amplitude size of the output cat.

{\it Conclusions.} In this work, we presented protocols for the deterministic preparation of squeezed cat states in optics. These protocols require a source of Fock states or even/odd parity states, PNRDs, and Gaussian measurements and operations. These elements are sufficient for universal fault-tolerant quantum computation using GKP error correction. It is also possible to do long-distance quantum communication by designing quantum repeaters using these encodings.

Feed-forward has the potential to improve the schemes considerably. For example, one may identify ``bad cases'' after a few rounds and adjust the beamsplitter ratios as required. A more sophisticated approach might target particular squeezing angles or squeezing strengths dependent on previous measurements.

The difficulty of our schemes is in part the resource state required. Fortunately, scheme II does not require a Fock state and the large-amplitude even/odd parity state can be built-up by combining even/odd parity elements (single-photons and squeezing) together in an optical network. Another possible method is to perform a non-destructive parity measurement on a large-amplitude state.

We hope that this work paves the way to future demonstrations in quantum information processing and error correction with squeezed cat and GKP codes.

\begin{acknowledgments}
This research was supported by the Australian Research Council Centre of Excellence for Quantum Computation and Communication Technology (Project No. CE170100012).
\end{acknowledgments}

\clearpage

\onecolumngrid

\renewcommand{\bibnumfmt}[1]{[S#1]}{}\renewcommand{\citenumfont}[1]{S#1}
\clearpage

\begin{center}
{\Large{SUPPLEMENTARY MATERIAL}}
\end{center}

$\;$

\renewcommand{\figurename}{Supplementary Figure}
\setcounter{figure}{0}

\section{Examples of single-mode states}

In Supplementary Figure~\ref{fig:Wigner_functions}, we plot Wigner functions (phase-space quasi-probability distributions) of important single-mode optical states used in this paper. From left to right in Supplementary Figure~\ref{fig:Wigner_functions}, we have the vacuum state $\ket{0}$, squeezed vacuum state $\hat{S}(r,\theta) \ket{0}$, Fock state $\ket{n}$ where $n$ is an integer, cat state $\propto \ket{\alpha}+\ket{-\alpha}$ where $\ket{\alpha} = \hat{D}(\alpha)\ket{0}$ are coherent states, squeezed cat state, and approximate GKP state.

\begin{figure}[H]
\centering
 \includegraphics[width=1\linewidth]{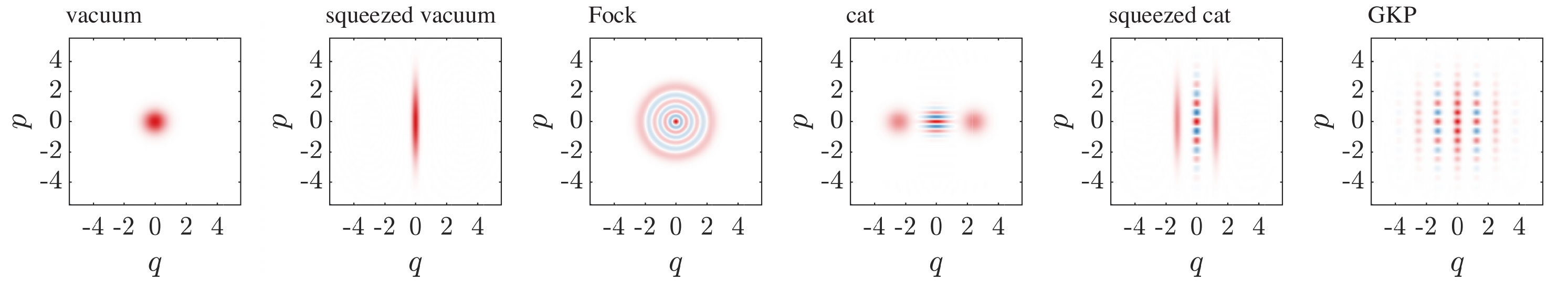}
 \caption{Wigner function plots of single-mode optical states with axes defined as $q = \text{Re}(\alpha)$ and $p=\text{Im}(\alpha)$ where $\alpha$ is the complex amplitude. From left to right we display the vacuum state $\ket{0}$, squeezed vacuum state $\hat{S}(r)\ket{0}$ with 10 dB of squeezing, Fock state $\ket{n}$ with $n=6$, two-component cat state $\propto \ket{\alpha}+\ket{-\alpha}$ where $\alpha = \sqrt{6}$, squeezed cat with amplitude $\alpha = \pm \sqrt{\pi/2}$ and 10 dB of squeezing, and physical GKP state $\ket{0_\text{GKP}^\Delta}$ with lattice spacing $\alpha = \sqrt{\pi/2}$ and 10 dB of squeezing ($\Delta = 0.316$).}\label{fig:Wigner_functions}
\end{figure}

\section{Preparing large-amplitude cat states non-deterministically using homodyne detection}\label{sec:homodyne_protocol}

Ref.~\cite{SUPP_Ourjoumtsev2007} showed how to prepare approximate squeezed cat states non-deterministically from a source of Fock states and homodyne detection. Consider the following circuit~\cite{SUPP_Ourjoumtsev2007}:
\begin{figure} [H]
\centering
 \includegraphics[width=0.3\linewidth]{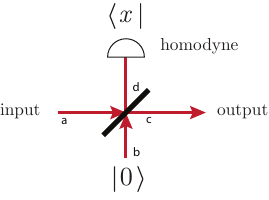}
\end{figure}

For an arbitrary input state with maximum $n$ photons $\ket{\psi_\text{in}} = \sum_{j=0}^n c_j \ket{j}$, the output state conditioned on homodyne measurement outcome $x$ is
\begin{align}
    (\bra{x} \otimes \mathbb{I}) B \ket{\psi}_{ab} &= e^{-x^2/2} \sum_{j=0}^n \frac{c_j}{\sqrt{j!}} \sum_{N=0}^{j} \frac{1}{\sqrt{\pi 2^N N!}}  H_N(x) {j \choose {N}} {\eta}^{N/2} ({1-\eta})^{(j-N)/2} \sqrt{N! (j-N)!} \ket{j-N}_{c}. 
\end{align}

Consider choosing the initial state to be a Fock state $\ket{n}$, as in Ref.~\cite{SUPP_Ourjoumtsev2007}. A success outcome occurs for near zero homodyne measurements $x\approx0$, where the produced output is a good approximation to a squeezed cat state. However, since failures can occur where $x \not\approx 0$, the scheme is non-deterministic for preparing cat states. 

Unfortunately, iterating the protocol does not allow the scheme to deterministically prepare cat states. To see this, consider taking the output state as the input state to another round, but this time choosing a random homodyne quadrature direction. The state deterministically evolves towards a displaced squeezed vacuum state since the evolving output state loses its two-fold rotation symmetry, and it also loses its non-Gaussianity. The squeezing coming from the initial Fock state decreases because of the beamsplitter and the uncertainty profile is limited by Heisenberg's uncertainty principle as it appears in optics (i.e., the uncertainty profile of a pure displaced squeezed vacuum state). That is, the output state after many rounds is approximately a displaced squeezed vacuum state and is no longer useful for preparing Bosonic code states.

By replacing the homodyne detector with squeezing and PNRD, iterating such a circuit can deterministically prepare approximate two-component squeezed cat states. This is our scheme I for preparing squeezed cat states. We introduce scheme II by moving the squeezing operator onto the other side of the beamsplitter so that the auxiliary vacuum state is squeezed. This is advantageous for the following reasons. First, it is easier to prepare squeezed vacuum states than doing inline squeezing. Second, adding photons to the evolving state each round increases the output cat-state amplitude.

\section{Supplementary material for scheme I}

\subsection{How to choose $\eta_j$ for equal splitting up of the light}

The iterative version of scheme I iteratively taps a little light off of the evolving cat state via a succession of beamsplitters with transmissivity $\eta_j$ where $j = 1,2,...,k$. The total transmissivity experienced by the light is
\begin{align}
    \eta_\text{total} = \prod_{j=1}^{k} \eta_j.\label{eq:total_transmissivity}
\end{align} 
It is most practical to set $\eta_j = \eta$. However, this means unequal amounts of light fall on each detector.

Now, we write down how to find the $\eta_j$ so that equal amounts of light fall on each detector. This would then match the $(k+1)$-splitter version shown in Figure 1 of the main text where the $(k+1)$-splitter splits the light up between the output and detected modes while further splitting the detected modes equally so that equal power falls on the detectors.

We define the amount of output reflected light after the $j$th beamsplitter at mode $j$ as
\begin{align}
    R_j &= (1-\eta_j)\eta_{j-1}\eta_{j-2}...\eta_1
\end{align}
and we enforce that $R_j = c$ where $c$ is a constant for all $j$. Then the total transmissivity is still given by~\cref{eq:total_transmissivity}.

As an example, consider $k=6$ and $\eta_{\text{total}}=1/2$, then solving the required equations we find that the beamsplitter transmissivities should be chosen as follows:
\begin{align}
    \eta_1 &= 0.9167\\
    \eta_2 &= 0.9091\\
    \eta_3 &=0.9000\\
    \eta_4 &=0.8889\\
    \eta_5 &=0.8750\\
    \eta_6 &=0.8571,
\end{align}
with total transmissivity
\begin{align}
    \prod_{j=1}^k \eta_j &= 1/2,
\end{align}
and our definition quantifying the amount of reflected light is $R_j = 0.0833$ which is constant for all $j$ as required.

We see that $\eta_j$ are roughly constant for all $j$, i.e., $\eta_j \approx \eta = \eta_\text{total}^{1/k}$ which is why in practice we can set $\eta = \eta_\text{total}^{1/k}$.

\subsection{Output state for scheme I}\label{sec:sqz_photon_measurement_protocol}

Consider the following circuit, where we switch the homodyne detector with squeezing and PNRD:
\begin{figure} [H]
\centering
 \includegraphics[width=0.3\linewidth]{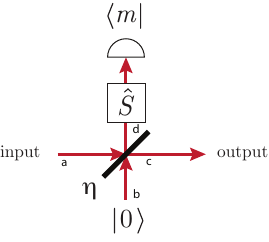}\label{fig:GKP_preparation2}
\end{figure}

Let the beamsplitter have transmissivity $\eta$, the squeezing parameter is $r$ in the $q$ direction (i.e., such a measurement is equivalent to $0$ outcome when homodyne $p$ as $r\to\infty$ when $m=0$), vacuum in mode $b$ $\ket{0}_b$, and detection of $m$ photons $\bra{m}_d$. Then given arbitrary input state in the Fock basis with maximum $n$ photons:
\begin{align}
    \ket{\psi_\text{in}}_a &= \sum_{j=0}^{n} c_j \ket{j}_a,
\end{align}
we calculate the output state of mode $c$ to be
\begin{multline}
   \ket{\psi_\text{out}}_c = \sum_{j=0}^{n} \sum_{N=0}^{j} \sum_l^{\min(N,m)}   \frac{c_j}{\cosh{(r)}^{(N+m+1)/2}}  \left(\frac{\sinh(r)}{2}\right)^{(N+m-2l)/2 } \\ \frac{(-1)^{(N-l)/2}N!}{l!(\frac{m-l}{2})!(\frac{N-l}{2})!}   \sqrt{\frac{{m!(j-N)! }}{{j!}}}
      { j \choose N} \eta^{N/2}(1-\eta)^{(j-N)/2} \ket{j-N}_c
\end{multline}
\begin{equation}
      \nonumber  l = \begin{cases}
     0,2,4,6 \dots \;\;\; N,m, \text{even}  \\
     1,3,5,7 \dots \;\;\; N,m, \text{odd}.  \\ 
      \end{cases} \label{eq:output}
\end{equation}

By choosing the initial input state to be $\ket{n}$, for large squeezing we see that the components of the output are located at the points given by the complex amplitude $\alpha$ with
\begin{align}
    \text{Re}(\alpha) &= \pm \sqrt{\frac{{-e^{-2r} |\alpha|^2 +b}}{{2\sinh(2r)}}} \\
    \text{Im}(\alpha) &= \pm \sqrt{|\alpha|^2-\text{Re}(\alpha)^2},
\end{align}
where we assume the real solutions. That is, $|\alpha|^2 = \text{Re}(\alpha)^2 + \text{Im}(\alpha)^2 $ and the four angles are given by $\tan{\phi} = \pm\text{Im}(\alpha)/\text{Re}(\alpha)$ and $\tan{\phi} = \pm \text{Im}(\alpha)/\text{Re}(\alpha) + \pi$.

If the curves do not intersect, then the components lie on the circle nearest the ellipse, i.e., if $\text{Re}(\alpha)=0$, then $\text{Im}(\alpha)=|\alpha|^2$, and if $\text{Im}(\alpha)=0$, then $\text{Re}(\alpha)=|\alpha|^2$. Clearly, the above expressions do not hold for low squeezing.

By choosing the initial input state to be $\ket{\psi_0}_a = \ket{n}$, after many rounds where the output becomes the input to the next round, choosing a random angle for the squeezing at each round, the state deterministically evolves towards $\ket{\psi_\text{many rounds}}_a \to \ket{\text{2-cat}}$.

\subsection{Squeezing followed by a photon-number measurement is a squeezed-cat state measurement near zero}\label{sec:sqz_m_equivalenet_to_sqz_cat}

Locally near the origin, a squeezed Fock state $\hat{S}_q\ket{m}$ appears like a superposition of quadrature eigenstates $\ket{q}+(-1)^m\ket{-q}$, depending on the parity of $m$. Since the input resource state for our protocol $\ket{n}$ has no more than $n$ photons, and all other auxiliary modes are vacuum states, it seems valid to consider the measurement locally near the origin.

Homodyne detection projects onto a quadrature eigenstate which is a Gaussian measurement that can prepare cat states non-deterministically. In contrast, projecting onto a superposition of quadrature eigenstates is a highly non-Gaussian measurement, projecting onto a superposition state, and can thus prepare cat states deterministically.

In this section we prove that $\hat{S}_q\ket{m} \sim \ket{q}+(-1)^m\ket{-q}$, depending on the parity of $m$, in some finite region $|p|\leq\epsilon$.

In this section we prove that a squeezed Fock state truncated to some finite region of the phase plane approaches a truncated superposition of quadrature eigenstates, i.e., a squeezed cat state. That is, we prove the following assuming squeezing in the $q$ quadrature ($r>0$):
\begin{align}
    \text{truncated}(\hat{S} \ket{m}) \sim \text{truncated}(\ket{e^{-r}\sqrt{m} ,r}+(-1)^m \ket{-e^{-r}\sqrt{m},r})\\
    \sim \text{truncated}(\ket{q=e^{-r}\sqrt{m}}+(-1)^m \ket{q=-e^{-r}\sqrt{m}}).
\end{align}

We first motivate the result geometrically noting that the Wigner function contour of $\hat{S}\ket{m}$ is an ellipse while for $\ket{q}+(-1)^m\ket{-q}$ it is two parallel lines. We show that in some finite region $|p|\leq\epsilon$ the area between the two curves goes to zero.

The formula for the ellipse for $\hat{S}\ket{m}$ is $\frac{x^2}{a^2} + \frac{y^2}{b^2} = 1$, where $a^2 = me^{2r}$ and $b^2 = me^{-2r}$. We are interested in the limit of large squeezing ($r\to\infty$) where $b>>a$. The area under the ellipse (between the positive region of the ellipse and the $y$ axis) in the finite region $y=|p|\leq\epsilon$ is
\begin{align}
    A &= \int_{-\epsilon}^{\epsilon} \frac{a}{b}\sqrt{b^2-y^2} \text{d}y\\
    &= \frac{a}{b} \left[ \epsilon \sqrt{b^2-\epsilon^2} + b^2 \tan^{-1} \left( \frac{\epsilon}{\sqrt{b^2-\epsilon^2}} \right) \right].
\end{align}
In the limit as $r \to \infty$, $b= me^{2r}\to\infty$, we have
\begin{align}
    \lim_{b\to\infty} A &= 2\epsilon a = 2 \epsilon \sqrt{m} e^{-r},
\end{align}
which is the area of a rectangle. That is, in the limit of large anti-squeezing, the ellipse in the region $|p|\leq\epsilon$ approaches parallel lines, and this motivates the fact that $\hat{S}_q\ket{m} \sim \ket{q}+(-1)^m\ket{-q}$, where $q = \sqrt{m}e^{-r}$.

We now show this result formally using the fidelity. The fidelity is
\begin{align}
    F &= \bra{m} \hat{S}(r)^\dagger (\ket{e^{-r}\sqrt{m},r}+(-1)^m \ket{-e^{-r}\sqrt{m},r})\\
    &= \bra{m} \hat{S}(r)^\dagger \hat{S}(r) (\ket{\sqrt{m}}+(-1)^m \ket{-\sqrt{m}})\\
    &= \bra{m} (\ket{\sqrt{m}}+(-1)^m \ket{-\sqrt{m}}),
\end{align}
where $\bra{m}$ is a Fock state bra and $\ket{\sqrt{m}}$ is a coherent state ket. $F<<1$ in general. But in a finite truncated region $|q|\leq\epsilon$ the fidelity approaches 1. To see this, we write Fock state $\ket{m}$ as a superposition of coherent states
\begin{align}
    \ket{m} &= \int_0^{2\pi} f(\phi)  \ket{\alpha e^{i\phi}}\; \text{d}\theta,
\end{align}
with $\alpha = \sqrt{m}$ and where
\begin{align}
    f(\theta) &= \frac{e^{\alpha^2/2}e^{-i m\theta} \sqrt{m!}}{2\pi \alpha^m}.
\end{align}
The truncated Fock state in the coherent state basis is then
\begin{align}
    \ket{m}_\text{trunc.} &\propto \int_0^{2\pi} f_\text{trunc.}(\theta)   \ket{\alpha e^{i\theta}} \; \text{d}\theta,
\end{align}

For each $\epsilon$, the angle subtended from the origin to $q= \pm \epsilon$ is
\begin{align}
    \phi = 2 \tan^{-1}\left(\frac{\epsilon}{\sqrt{m}}\right).
\end{align}
For a fixed $p=\sqrt{m}e^{-r}$ value, there is a relationship between $r$ and $m$: $\sqrt{m} = p e^{r}$. That is, large squeezing $r>>0$ requires large $m$ since the Fock state is squeezed, thus,
\begin{align}
    \lim_{r\to \infty} \phi = \lim_{m\to\infty} \phi = 0.
\end{align}
Then, the only coherent-state components of our anti-squeezed region lie precisely on the $q$ axis, and $f_\text{trunc.}(\theta)$ then approaches only two values at $\theta =0$ and $\theta = \pi$ so that
\begin{align}
    \lim_{r \to \infty} \ket{m}_\text{trunc.} &= \mathcal{N} ( \ket{\alpha}\pm\ket{-\alpha}),
\end{align}
so that
\begin{align}
    \lim_{r\to\infty} F_\text{trunc.} &= \lim_{r\to\infty} \bra{m}_\text{trunc.} 
 \mathcal{N} (\ket{\alpha}+\ket{-\alpha})\\
 &= \mathcal{N}^2 (\bra{\alpha}\pm\bra{-\alpha})( (\ket{\alpha}\pm\ket{-\alpha}) = 1.
\end{align}

Geometrically, what happens is that $\epsilon$ effectively goes to zero as the state is infinitely squeezed.

\subsection{A 4-cat projective measurement}\label{sec:het_cat_measurement}

To obtain the same noise profile on the components of the output cat state for any measurement outcome, it is convenient to perform the measurement in two orthogonal squeezing directions, $\hat{q}$ and $\hat{p}$, (like heterodyne detection) as follows:
\begin{figure}[H]
\centering
 \includegraphics[width=0.5\linewidth]{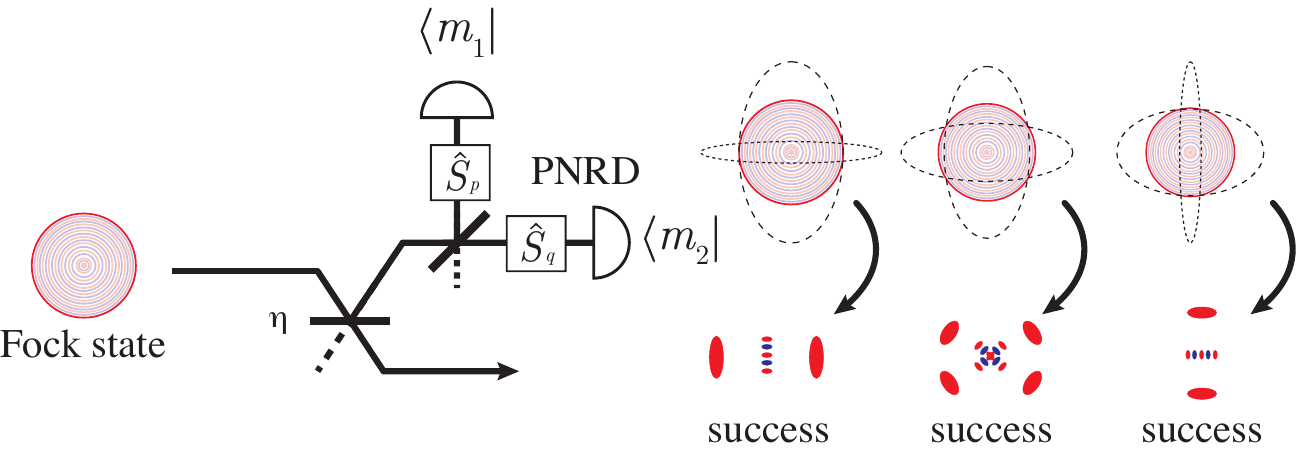}
\end{figure}
\noindent
The measurement projects onto $ (\bra{0}\otimes \mathbb{I}) \hat{B}(1/2) \hat{S_p} \ket{m_2} \hat{S_q} \ket{m_1}$ which is equivalent to projecting onto a four-component cat state $\sum_{k=0}^3 \ket{\gamma_k}$ in the limit of large squeezing. We now prove this.

In this section, we prove that the following measurement projects onto a 4-cat:
\begin{figure}[H]
\centering
\includegraphics[width=0.15\linewidth]{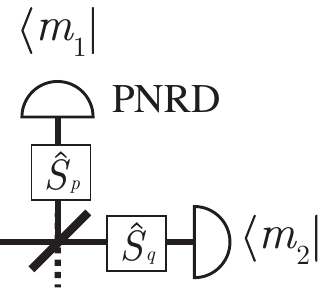}
\end{figure}
That is,
\begin{align}
    \ket{\psi_\text{out}} &= \bra{m_1}\hat{S}_q \bra{m_2}\hat{S}_p \hat{B}(\eta=0.5) \ket{\psi_\text{in}}\ket{0}\\
    &= \braket{\text{4-cat}}{\psi_\text{in}}.
\end{align}

First, due to the time-reversal property of quantum mechanics, we can reverse the circuit:
\begin{figure}[H]
\centering
\includegraphics[width=0.15\linewidth]{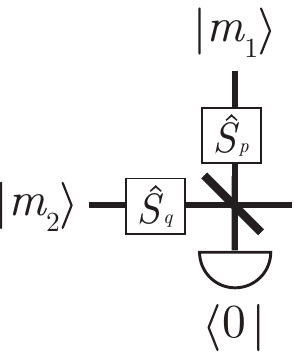}
\end{figure}
that is, we want to show that
\begin{align}
    (\mathbb{I} \otimes \bra{0}) \hat{B}(\eta=0.5) \hat{S}_q \ket{m_2} \hat{S}_p \ket{m_1} &= \ket{\text{4-cat}}.
\end{align}

We make use of the following definition for the quadrature operators:
\begin{align}
    \hat{q_j} &\equiv \hat{a}_j + \hat{a}_j^\dagger\\
    \hat{p_j} &\equiv i(- \hat{a}_j + \hat{a}_j^\dagger),
\end{align}
for each mode $j$.

The balanced beamsplitter transforms the quadrature operators in the following way:
\begin{align}
    q_1 \to (q_1+q_2)/\sqrt{2}\\
    p_1 \to (p_1+p_2)/\sqrt{2}\\
    q_2 \to (-q_1+q_2)/\sqrt{2}\\
    p_2 \to (-p_1+p_2)/\sqrt{2}.
\end{align}
Projecting mode two onto zero means $q_1=q_2$ and $p_1 = p_2$ then
\begin{align}
    q_1 \to \sqrt{2} q_1 \\
    p_1 \to \sqrt{2} p_1 .
\end{align}
Owing to $S_p \ket{m_1}\sim \ket{p}+(-1)^{m_1}\ket{-p}$ and $S_q \ket{m_2}\sim \ket{q}+(-1)^{m_2}\ket{-q}$ near the truncated Hilbert space for large squeezing, the output state is
\begin{multline}
    (\mathbb{I} \otimes \bra{0}) \hat{B} \hat{S}_q \ket{m_2} \hat{S}_p \ket{m_1} =
   \ket{(\sqrt{m_2}+i\sqrt{m_1})e^{-|r|}\sqrt{2}} + (-1)^{m_1}\ket{(\sqrt{m_2}-i\sqrt{m_1})e^{-|r|}\sqrt{2}} \\  +  (-1)^{m_2}\ket{(-\sqrt{m_2}+i\sqrt{m_1})e^{-|r|}\sqrt{2}}+ (-1)^{m_1+m_2}\ket{(-\sqrt{m_2}-i\sqrt{m_1})e^{-|r|}\sqrt{2}},\label{eq:4-cat}
\end{multline}
where $q = 2\sqrt{2m_2}e^{-|r|}$ and $p=2\sqrt{2m_2}e^{-|r|}$.

Thus, the measurement projects onto a 4-cat 
\begin{align}
    (\mathbb{I} \otimes \bra{0}) \hat{B} \hat{S}_q \ket{m_2} \hat{S}_p \ket{m_1} &\sim \ket{\text{4-cat}} = \mathcal{N}_\text{4-cat} \sum_{k=0}^3 \pm \ket{\alpha e^{i\theta_k}},
\end{align}
where $\alpha$ is here assumed to be real with amplitude $|\alpha| = \sqrt{2(m_1+m_2)}e^{-|r|}$ at phase angles $\theta_k = \{\theta,-\theta,\theta+\pi,-\theta+\pi\}$ where $\theta = \tan^{-1}\sqrt{m_1/m_2}$, and the phases between the components are determined from the parity of $m_1$ and $m_2$, shown explicitly in~\cref{eq:4-cat}.

\subsection{Proof that the first round of scheme I deterministically prepares 4-cats}
\label{sec:proof_cat_prep}

We will show that
\begin{align}
    (\mathbb{I}  \otimes \bra{\text{4-cat}}) B(\eta) \ket{n}\ket{0} = \ket{\text{4-cat}}.
\end{align}

Write $\ket{n}$ in the Fock basis~\cite{SUPP_PhysRevA.48.2213}:
\begin{align}
    \ket{n} &= \int_0^{2\pi} \text{d}\theta \; f(\theta) \ket{\alpha e^{i\theta}},
\end{align}
where
\begin{align}
    f(\theta) &= \frac{e^{\alpha^2/2}e^{-i n\theta} \sqrt{n!}}{2\pi \alpha^n},
\end{align}
and it is convenient to choose $\alpha = \sqrt{n}$.

The initial state before the beamsplitter is
\begin{align}
     \ket{n}\ket{0}&=\int_0^{2\pi} \text{d}\theta \; f(\theta) \ket{\alpha e^{i\theta}}\ket{0}.
\end{align}

After the beamsplitter
\begin{align}
     B(\eta) \ket{n}\ket{0} &= \int_0^{2\pi} \text{d}\theta \; f(\theta) \ket{\sqrt{\eta} \alpha e^{i\theta}} \ket{\sqrt{1-\eta} \alpha e^{i\theta}},
\end{align}
where we have corrected the phase difference between the transmitted and reflected beams. The two modes are entangled in phase.

Finally, after the measurement
\begin{align}
     (\mathbb{I}  \otimes \bra{\text{4-cat}}) \int_0^{2\pi} \text{d}\theta \; f(\theta) \ket{\sqrt{\eta} \alpha e^{i\theta}} \ket{\sqrt{1-\eta} \alpha e^{i\theta}} &= (\mathbb{I}  \otimes \mathcal{N}_\text{4-cat} \sum_{k=0}^3 \pm \bra{\beta e^{i\theta_k}}) \int_0^{2\pi} \text{d}\theta \; f(\theta) \ket{\sqrt{\eta} \alpha e^{i\theta}} \ket{\sqrt{1-\eta} \alpha e^{i\theta}}\\
    &=    \int_0^{2\pi} \text{d}\theta \; f(\theta) \mathcal{N}_\text{4-cat} \sum_{k=0}^3 \pm \braket{\beta e^{i\theta_k}}{\sqrt{1-\eta} \alpha e^{i\theta}}  \ket{\sqrt{\eta} \alpha e^{i\theta}},
\end{align}
where $\braket{\beta e^{i\theta_k}}{\sqrt{1-\eta} \alpha e^{i\theta}}  = e^{- \frac{\alpha^2 (1-\eta)}{2}} e^{- \frac{\beta^2}{2}} e^{\alpha \beta \sqrt{1-\eta} e^{i\theta} e^{-i\theta_k}}$, i.e.,
\begin{align}
    (\mathbb{I}  \otimes \bra{\text{4-cat}}) B(\eta) \ket{n}\ket{0} &= \int_0^{2\pi} \text{d}\theta \; g(\theta) \ket{\alpha \sqrt{\eta} e^{i\theta}},
\end{align}
where
\begin{align}
    g(\theta) =  f_\theta \mathcal{N}_\text{4-cat} \left(\sum_{k=0}^3 \pm e^{- \frac{\alpha^2 (1-\eta)}{2}} e^{- \frac{\beta^2}{2}} e^{\alpha \beta \sqrt{1-\eta} e^{i\theta} e^{-i\theta_k}} \right),
\end{align}
where $\alpha = \sqrt{n}$ and $|\beta| = \sqrt{2(m_1+m_2)}e^{-|r|}$ with $\theta_k = \{\theta_1,-\theta_1,\theta_1+\pi,-\theta_1+\pi\}$ where $\theta_1 = \tan^{-1}\sqrt{m_1/m_2}$.

That is, the probability distribution consists of sharp peaks located at $\theta_k$. We approximately have a superposition of four squeezed states with approximately 3 dB of squeezing for $\eta=1/2$:
\begin{align}
    (\mathbb{I}  \otimes \bra{\text{4-cat}}) B(\eta) \ket{n}\ket{0} &\propto \sum_{k=0}^3 \pm \ket{\sqrt{\eta n} e^{i\theta_k}, r=-\frac{1}{2}\ln({1-\eta})},
\end{align}
where $\theta_k = \{\theta,-\theta,\theta+\pi,-\theta+\pi\}$. Thus, we deterministically prepare a 4-cat with vanishing phase variance for large $n$, meaning we can certainly remove two of the components by doing the protocol again, rotating first by some angle not equal to multiples of $\pi/2$. In this way, we can deterministically remove one of the pairs of components from the 4-cat.

\subsection{Transforming Output States into Unsqueezed Cat States or Approximate GKP States}\label{sec:transforming_output}

Recall that the output states are cat states, with components that already have some squeezing. This output state is
\begin{align}
    \ket{\mathcal{C}_{\alpha,r}^\pm} \equiv \mathcal{N} ( \ket{\alpha,r} \pm \ket{-\alpha,r}) = \mathcal{N} [ \hat{D}(\alpha)\hat{S}(r) \pm \hat{D}(-\alpha)\hat{S}(r) ] \ket{0}. \label{eq:catds}
\end{align}
In particular, notice that the squeezing is first and then the displacement is applied. We found that scheme I generates output cat states with approximately $|\alpha|=\sqrt{n\eta^k}$ amplitude and $r=-0.5\ln(1-\eta^k)$ squeezing. 

Here we will consider two different situations: (a) The amount of anti-squeezing required to turn the output state into an unsqueezed cat state. We will also determine how the amplitude of this unsqueezed cat state scales with the protocol parameters. This will be helpful for users who want to use our schemes for standard cat state generation. (b) The amount of extra squeezing required to squeeze the output state onto the correct grid position of $\pm\sqrt{\pi/2}$ required for approximate GKP states. We will also figure out the final amount of squeezing on the components $\Delta$, as a function of the protocol parameters.

We will use the following mathematical identity
\begin{align}
    \hat{D}(\alpha)\hat{S}(re^{i\phi}) = \hat{S}(re^{i\phi})\hat{D}(\gamma), \quad \gamma = \alpha \cosh(r) + \alpha^* e^{i\phi} \sinh(r),
\end{align}
which is given in Ref.~\cite{SUPP_gong1990expansion}. However, we can simplify this expression in our case, considering the squeezing is always in the amplitude direction. Without loss of generality, if we consider rotating our output cat state such that the components aligns with the horizontal $q$-axis (i.e., $\alpha=\sqrt{n\eta^k}$ is real), then squeezing direction is also horizontal (i.e., $\phi=0$). In this case $\gamma = \alpha e^r$, hence the identity is greatly simplified to
\begin{align}
    \hat{D}(\alpha)\hat{S}(r) = \hat{S}(r)\hat{D}(\alpha e^r). \label{eq:dstosd}
\end{align}
We will use this method of switching the order of operations (and its inverse) to derive the expressions we want. 

\subsubsection{Creating Unsqueezed Cat States} 

Consider switching the operations in \cref{eq:catds} using \cref{eq:dstosd} as follows
\begin{align}
    \ket{\mathcal{C}_{\alpha,r}^\pm} &= \mathcal{N} [ \hat{D}(\alpha)\hat{S}(r) \pm \hat{D}(-\alpha)\hat{S}(r) ] \ket{0} \\ 
    &= \mathcal{N} \hat{S}(r) [ \hat{D}(\alpha e^r ) \pm \hat{D}(-\alpha e^r) ] \ket{0} \\ 
    &= \hat{S}(r) \ket{\mathcal{C}_{\alpha e^r,0}^\pm}. 
\end{align}
This shows we can turn our output state into an unsqueezed cat state $\ket{\mathcal{C}_{\alpha e^r,0}^\pm}$ by applying $\hat{S}^\dagger(r) = \hat{S}(-r)$ squeezing operator. Note that $-r = 0.5\ln(1-\eta^k)$ corresponds to anti-squeezing, since $\eta^k \in [0,1]$. Furthermore, we can see that the unsqueezed cat state will have an amplitude of 
\begin{align}
    \alpha \exp(r) = \sqrt{n\eta^k} \exp(-0.5\ln(1-\eta^k)) = \sqrt{\frac{n\eta^k}{1-\eta^k}}. 
\end{align}
We can see that as $\eta^k$ approaches zero (i.e. the beamsplitter becomes fully reflective), the amplitude becomes zero. This makes sense as the output state should be vacuum in this limit. Note that as $\eta^k$ approaches one, the amplitude becomes large; however in this situation the output state is not a cat state (rather it is a Fock state), therefore this expression is not valid in this limit. 

\subsubsection{Creating Approximate GKP States}

Let us consider applying extra squeezing $\hat{S}(z)$ to the output cat state, so that we can get the approximate GKP states. Using the inverse of \cref{eq:dstosd} we can show the effect of this extra squeezing as
\begin{align}
    \hat{S}(z)\ket{\mathcal{C}_{\alpha e^r,0}^\pm} &= \mathcal{N} [ \hat{S}(z) \hat{D}(\alpha)\hat{S}(r) \pm \hat{S}(z)\hat{D}(-\alpha)\hat{S}(r) ] \ket{0} \\ 
    &= \mathcal{N} [ \hat{D}(\alpha e^{-z})  \hat{S}(z) \hat{S}(r) \pm \hat{D}(-\alpha e^{-z}) \hat{S}(z)\hat{S}(r) ] \ket{0} \\
    &=\mathcal{N} [ \hat{D}(\alpha e^{-z}) \hat{S}(z+r) \pm \hat{D}(-\alpha e^{-z}) \hat{S}(z+r)  ] \ket{0}\\ 
    &= \ket{\mathcal{C}_{\alpha e^{-z},z+r}^\pm}. \label{eq:sqcat}
\end{align}
Recall for approximate GKP states we want to align the components on the correct grid positions, which is $\pm\sqrt{\pi/2}$. This displacement requirement for \cref{eq:sqcat} means the magnitude of the extra squeezing should be
\begin{align}
    \alpha e^{-z} = \sqrt{\frac{\pi}{2}} \quad \Rightarrow \quad z = - \ln \left( \frac{1}{\alpha} \sqrt{\frac{\pi}{2}}  \right) = -\ln \left( \sqrt{\frac{\pi}{2n\eta^k}}  \right). 
\end{align}
This allows us to now determine the final squeezing associated with the components as
\begin{align}
    z+r =  -\ln \left( \sqrt{\frac{\pi}{2n\eta^k}} \right) - 0.5\ln(1-\eta^k) = - \ln \left(  \sqrt{\frac{\pi(1-\eta^k)}{2n\eta^k}}  \right). 
\end{align}
Hence we know after applying this extra squeezing, the squeezing parameter associated with the approximate GKP state is 
\begin{align}
    \Delta = \sqrt{\frac{\pi(1-\eta^k)}{2n\eta^k}}. \label{eq:deltagkp} 
\end{align}

\section{Supplementary material for scheme II}

\subsection{Approximation to the phase probability distribution evolution}\label{sec:approximation_phase_distribution}

A good approximation for the evolving probability distribution for scheme II is
\begin{align}
    P_\text{out} &= \frac{\prod_j P_j}{\int_0^{2\pi} \prod_j P_j \;\text{d}\phi},
\end{align}
where the phase probability distribution $P_j$ depends on outcome $m_j$ at the PNRD and the input squeezing direction $\theta_j$, where $j$ is the round number. $P_j$ are approximately cosine functions when the total number of rounds is large $(k\gg1)$. In this case, a good approximation is $P_{j}(m_j=\text{even})\approx 1$ and $P_{j}(m_j=\text{odd})\approx-\cos(2\phi-\phi_j/2)+1$, where $\phi_j\in(0,\pi]$ are effective phase measurement outcomes. That is, at round $j$, the phase probability distribution changes significantly if the outcome at the PNRD is odd and the evolving state changes parity. 

Simulating $P_\text{out}$ for many iterations we always eventually get two strong peaks at $\phi_\text{out} \in(0,\pi]$ and $\phi_\text{out}+\pi$, that is, we deterministically evolve towards a two-component cat state. To see this, take the effective phase measurement outcomes $\phi_j$ to be totally random and start with a constant phase probability distribution $P_0 = \text{constant}$.

\subsection{Using non-determinism to decrease the required Fock resource state}

The trade-off between the success probability and the obtainable size of the cats using a source of squeezed Fock states is illustrated in Supplementary Figure~\ref{fig:squeezed_Fock_number}.

\begin{figure}
\centering
\includegraphics[width=0.4\linewidth]{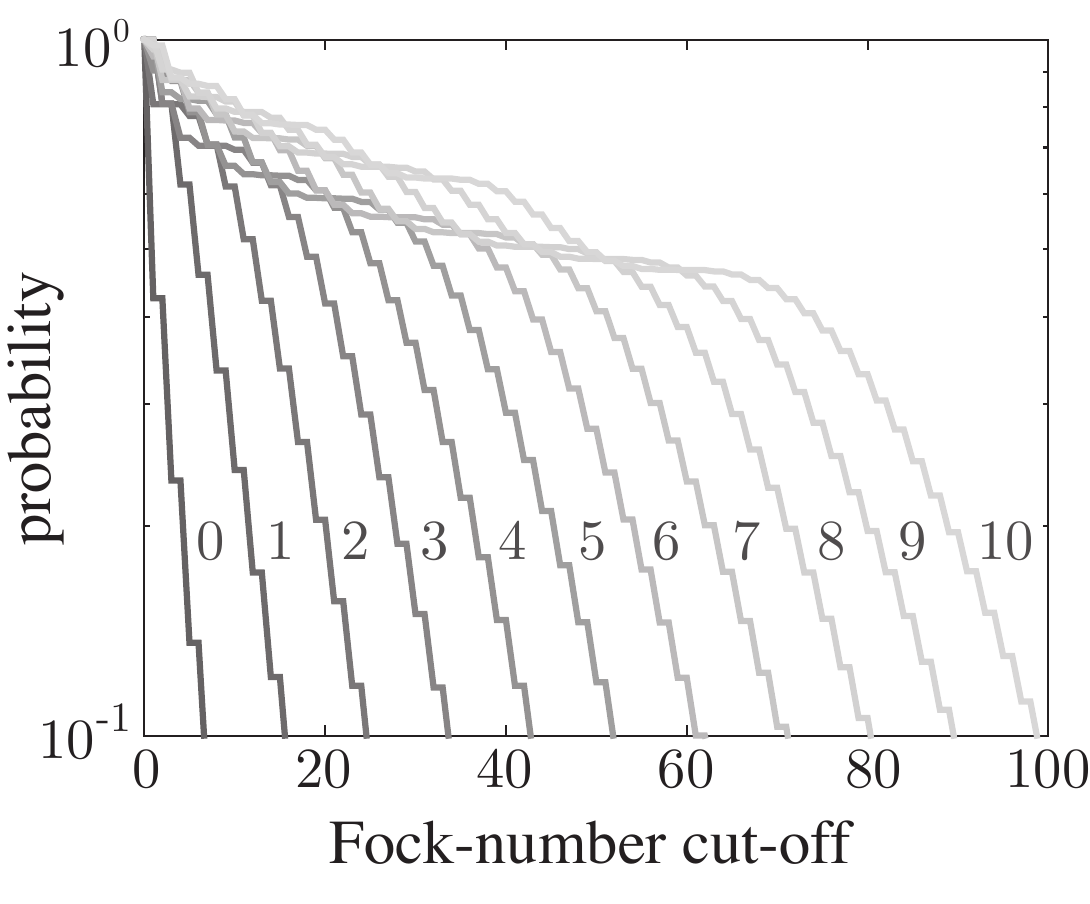}
  \caption{Probability that a squeezed Fock state $\hat{S}(r)\ket{n}$ with 10 dB of squeezing has at least $n_{\text{cut-off}}$ number of photons given on the horizontal axis, that is, letting $\hat{S}(r)\ket{n} = \sum_{k=0}^{\infty} c_k \ket{k}$ then probability is $P = \sum_{k=n_{\text{cut-off}}}^{\infty} |c_k|^2$. This approximates the probability for achieving a minimum desired squared cat-amplitude size ($n_\text{cut-off} \lesssim |\alpha_\text{desired cat}|^2$) and the achievable minimum squeezing for GKP state preparation $ 10\log_{10}(n_\text{cut-off})$ dB. Each curve is for a different squeezed Fock state with $n$ labelled on the figure. The required $n$ can be much less than $|\alpha_\text{desired cat}|^2$ because of squeezing and post-selection.}\label{fig:squeezed_Fock_number}
\end{figure}

The mean photon number of squeezed Fock states $\ket{n}$ as a function of the squeezing parameter $r$ is shown in Supplementary Figure~\ref{fig:mean_photon_number}.

\begin{figure}
\centering
\includegraphics[width=0.8\linewidth]{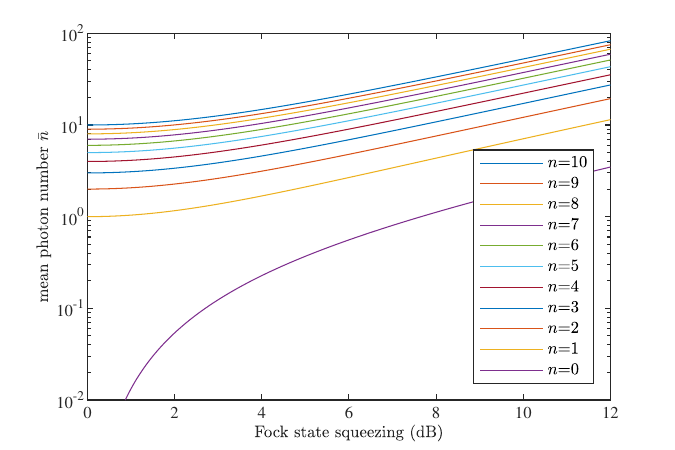}
  \caption{Mean photon number of a squeezed Fock state $\ket{n}$ as a function of the amount of Fock state squeezing (dB) for a given $n$. This plot illustrates the average number of photons available at the input to scheme II for cat-state preparation leading to larger prepared cats as a function of the Fock state squeezing. Remarkably, this plot indicates a great advantage to using a squeezed single photon input rather than a squeezed vacuum input.}\label{fig:mean_photon_number}
\end{figure}

\section{Experimental imperfections}\label{sec:imperfections_details}

In this section, we model some imperfections with noisy quantum channels. Consider placing noise into the circuit for scheme I as marked in the following, at either on the transmitted evolving cat state (a), before the squeezing (b), or before the detector (c), as follows

\begin{figure}[H]
\centering
\includegraphics[width=0.35\linewidth]{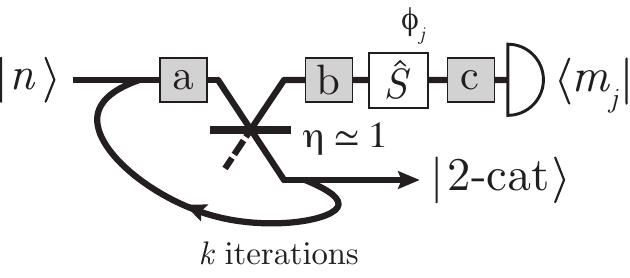}
\end{figure}

We find that the scheme is more tolerant to noise on the reflected beam than noise on the transmitted beam since the transmitted beam is the evolving cat state.

\subsection{Losses}

We model imperfect detectors with efficiency $\epsilon\in[0,1]$, by placing a pure-loss channel immediately before the detector with transmissivity $\epsilon$, followed by an ideal detector. Similarly, we can also model an inefficient squeezing operation by placing a pure-loss channel before the squeezing. Our scheme is tolerant to losses on the reflective beam when the transmissivity is close to one $\eta\simeq1$, since not much of the signal is reflected. It is important to minimise the losses on the transmitted mode of the evolving cat state during the protocol to preserve the interference fringes of the cat state.

Example Wigner functions of the output states modelling inefficient squeezing and detectors is plotted in Supplementary Figure~\ref{fig:losses}, that is, with lossy channels marked at locations b and c. The total transmissivity of the lossy channels is shown in the figure.

\begin{figure}
\centering
\subfigure[]{\label{fig:losses_a}\includegraphics[width=0.245\linewidth]{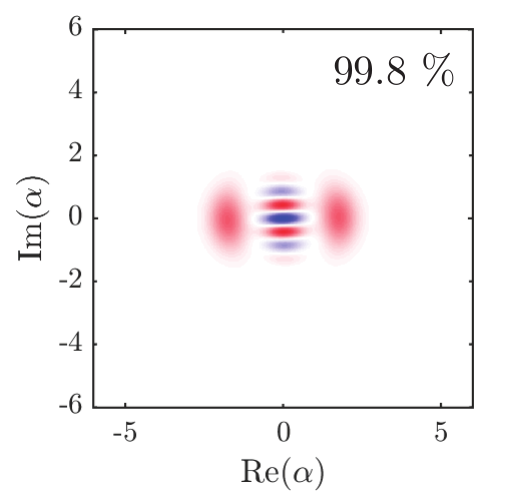}}
\subfigure[]{\label{fig:losses_b}\includegraphics[width=0.245\linewidth]{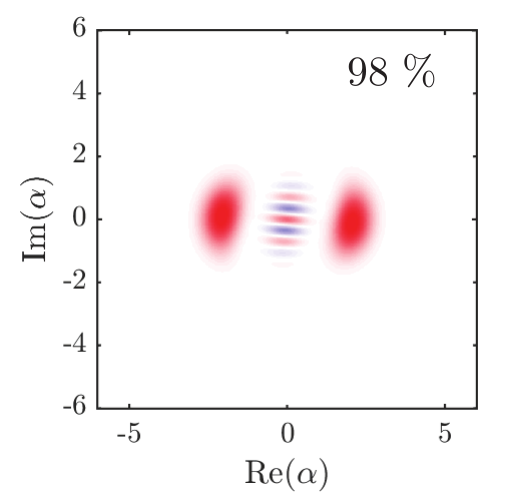}}
\subfigure[]{\label{fig:losses_c}\includegraphics[width=0.245\linewidth]{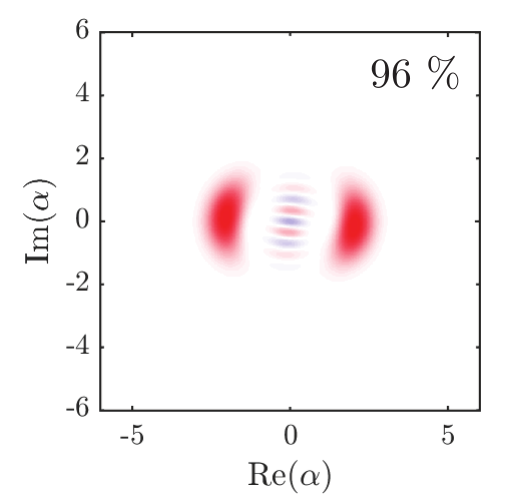}}
\subfigure[]{\label{fig:losses_d}\includegraphics[width=0.245\linewidth]{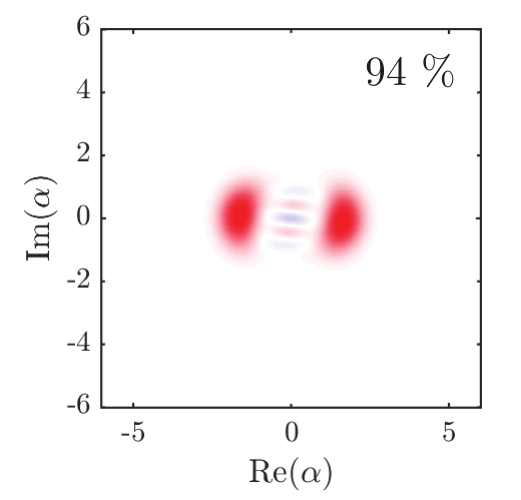}}
 \caption{Wigner functions of the output state of the iterative cat-preparation protocol, modelling inefficient squeezing and detectors with equal loss before and after the squeezing (i.e., at locations b and c), with total transmissivity as shown. The initial Fock state is $\ket{n=20}$ with $r = -0.5$ amount of 
 inline squeezing. The beamsplitter transmissivity is $\eta = 0.85$ with 10 iterations.}\label{fig:losses}
\end{figure}

\subsection{Dephasing}

Dephasing noise is modelled as random rotations in phase space. The Kraus operator representation is given in eq. F12 of~\cite{SUPP_PhysRevX.10.011058}. Numerical results for including dephasing everywhere in the circuit are shown in Supplementary Figure~\ref{fig:dephasing}.

\begin{figure}
\centering
\subfigure[]{\label{fig:dephasing_a}\includegraphics[width=0.245\linewidth]{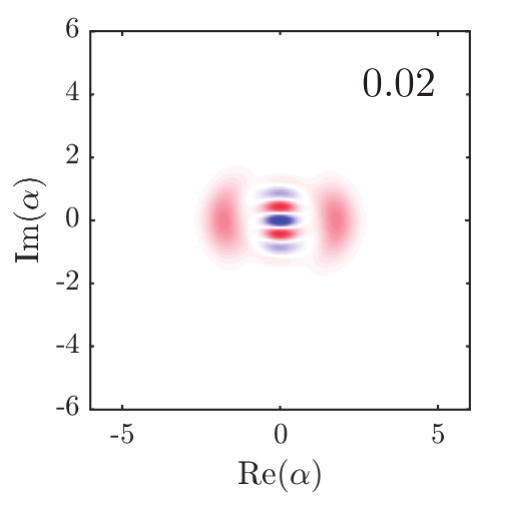}}
\subfigure[]{\label{fig:dephasing_b}\includegraphics[width=0.245\linewidth]{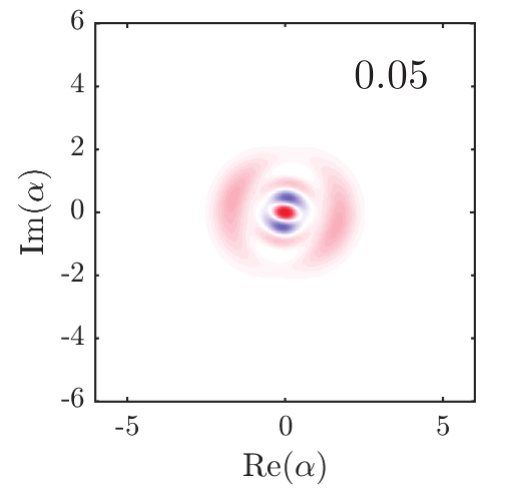}}
\subfigure[]{\label{fig:dephasing_c}\includegraphics[width=0.245\linewidth]{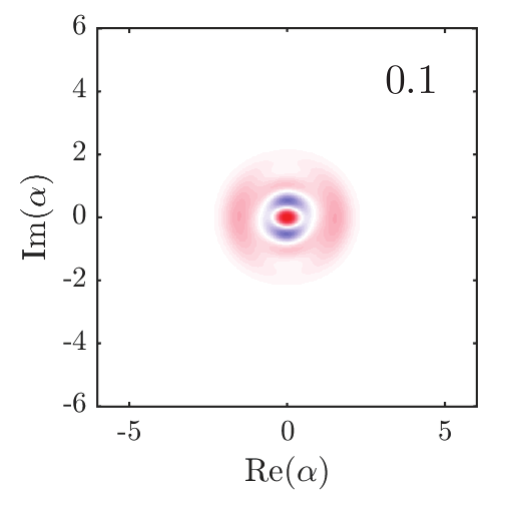}}
\subfigure[]{\label{fig:dephasing_d}\includegraphics[width=0.245\linewidth]{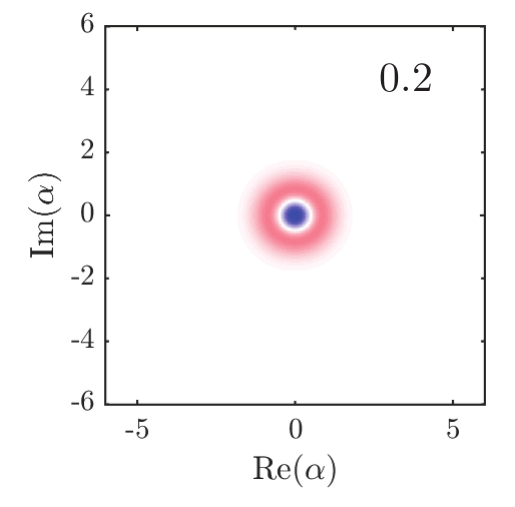}}
\caption{Same as Supplementary Figure~\ref{fig:losses}, but instead of loss, we consider dephasing noise parameterised by $\epsilon_\phi$ (as shown on the figures) at all locations a, b, and c.}\label{fig:dephasing}
\end{figure}

\section{Error correction performance}\label{sec:GKP_error_correction}

In this section we verify that the prepared GKP states using our schemes can be used for error correction. In particular, we define an average optimal entanglement fidelity which shows how well the states perform against a specific error model. In this case, we assume a pure-loss channel.

\subsection{Average optimal entanglement fidelity}

In this section, we calculate the average optimal entanglement fidelity using our schemes. This is a figure of merit for GKP error correction against noise.

Consider preparing two squeezed cat states using our scheme, and then combining them to make a GKP state. This is one level of breeding, which produce GKP states $\ket{\psi_\text{bred GKP}}$ with three peaks. These GKP states make up our logical code with $\ket{0_\text{L}} = \ket{\psi_\text{bred GKP}}$ and $\ket{1_\text{L}} = \hat{D}(\sqrt{\pi/2})\ket{\psi_\text{bred GKP}}$. We encode one half of a qubit Bell pair into the code, apply loss to the encoded qubit (which is the dominant source of noise in optics), apply an optimal error correction recovery, then decode back to the qubit space, and compute the fidelity of the output entangled pair with the initial Bell pair. This defines our entanglement fidelity for a given prepared code against loss. We use a semi-definite program (SDP) using CVX~\cite{SUPP_cvx} in MATLAB~\cite{SUPP_MATLAB} to obtain the optimal entanglement fidelity, as used in Ref.~\cite{SUPP_PhysRevA.97.032346}. We compute this for several runs of the experiment to obtain an average optimal entanglement fidelity.

That is, we combine two random cats to prepare a logical zero GKP state and we combine a different random pair of cats to prepare logical one.

Numerical results are plotted in Supplementary Figure~\ref{fig:entanglement_fidelity} as a function of initial photon number size $n$ using scheme II (purple). We average over 10 complete simulations of the GKP breeding experiment. We also show the break-even point for error correction given by the trivial code $\ket{0_\text{L}}\equiv\ket{0}$ and $\ket{1_\text{L}}\equiv\ket{1}$ (black) as well as a trivial squeezed-coherent state code (grey) given by $\ket{0_\text{L}}\equiv \ket{\alpha_0,r}$ and $\ket{1_\text{L}}\equiv\ket{\alpha_1,r}$, where $\alpha_0=0$ and $\alpha_1 = \sqrt{\pi/2}$ and the amount of squeezing is 10 dB.

This plot shows how the cat and GKP preparation schemes are robust to loss, since the loss is corrected using GKP error correction. We place more details about experimental imperfections in~\cref{sec:imperfections_details}.

Note that these results employ an approximation since the Hilbert space is numerically truncated, also we have not averaged over the full possibility of outcomes of which there are many. In any case, according to Supplementary Figure~\ref{fig:entanglement_fidelity}, the GKP state preparation scheme seems to become useful for error correction when $n \gtrsim 4$. It would be very interesting to explore the limitation and performance of our GKP state-preparation schemes for quantum error correction in future works.

\begin{figure}
\centering
 \includegraphics[width=0.5\linewidth]{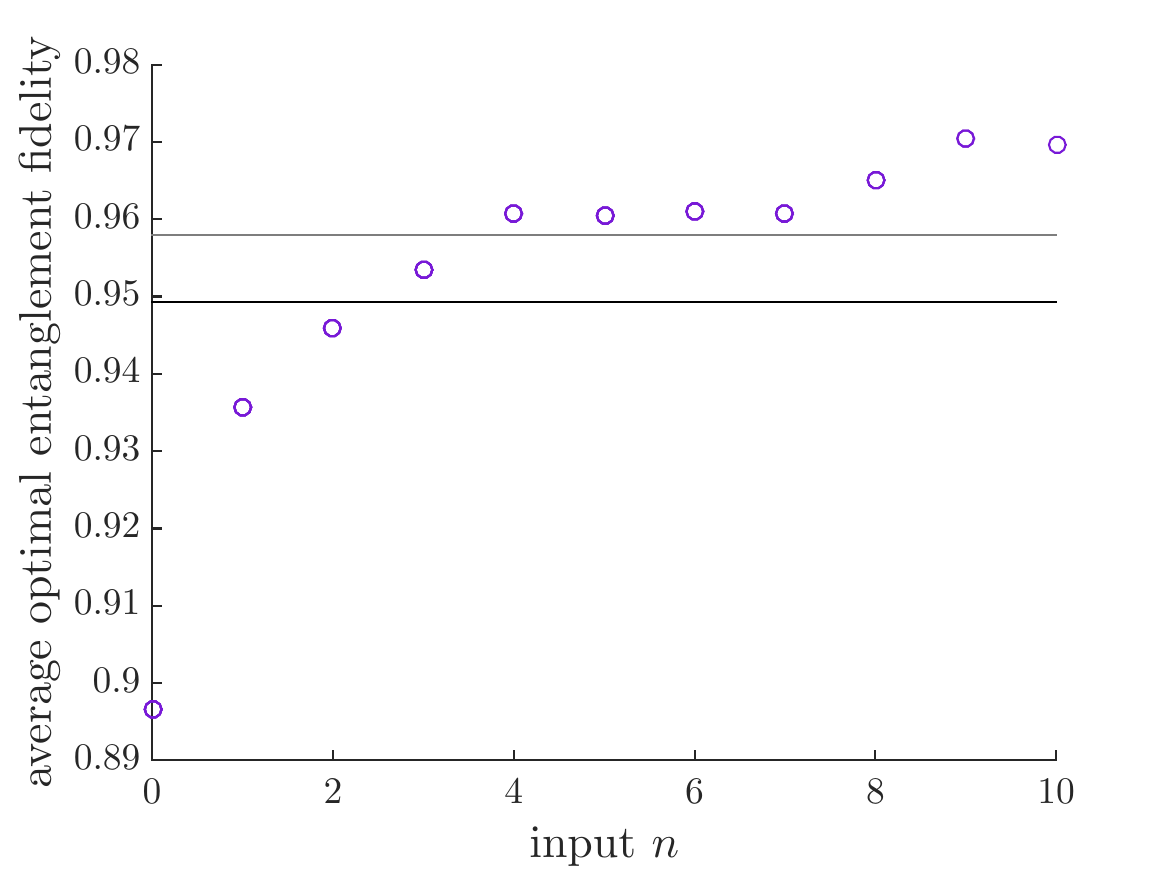}
 \caption{Numerical simulation results for GKP state preparation using scheme II assuming a source of input Fock states are available for cat-state preparation and one rounds of GKP breeding from the squeezed cat states. We plot the average optimal entanglement fidelity as a function of initial photon number $n$, assuming that the encoded entangled Bell pair is acted on by a  pure-loss error channel with transmissivity $\eta = 0.9$, followed by an optimal recovery operation. The results are averaged over 10 runs of the experiment (giving rough results only). Squeezed cat states are deterministically prepared and randomly bred into GKP states using just one round of breeding, consuming two squeezed cat states per logical GKP state. The break-even point for error correction is shown via the black curve for the trivial code ($\ket{0_\text{L}}\equiv\ket{0}$ and $\ket{1_\text{L}}\equiv\ket{1}$) and the grey curve for the squeezed-coherent state code ($\ket{0_\text{L}}\equiv \hat{S}(r)\ket{0}$ and $\ket{1_\text{L}} \equiv \hat{D}(\sqrt{\pi/2})\hat{S}(r)\ket{0}$) with 10 dB of squeezing. We proved analytically, and this plot confirms numerically, that GKP states prepared using our schemes can achieve the break-even point for error correction if the amplitude of the input states are sufficiently large.}\label{fig:entanglement_fidelity}
\end{figure}

\end{document}